\documentclass[bibyear]{aa}  

\usepackage{graphicx}
\usepackage{txfonts}
\usepackage{natbib}
\bibpunct{(}{)}{;}{a}{}{,} 
\usepackage{booktabs} 
\usepackage{natbib,twoopt}
\usepackage[breaklinks=true]{hyperref} 
\bibpunct{(}{)}{;}{a}{}{,}             
\makeatletter
  \newcommandtwoopt{\citeads}[3][][]{\href{http://adsabs.harvard.edu/abs/#3}%
    {\def\hyper@linkstart##1##2{}%
     \let\hyper@linkend\@empty\citealp[#1][#2]{#3}}}
  \newcommandtwoopt{\citepads}[3][][]{\href{http://adsabs.harvard.edu/abs/#3}%
    {\def\hyper@linkstart##1##2{}%
     \let\hyper@linkend\@empty\citep[#1][#2]{#3}}}
  \newcommandtwoopt{\citetads}[3][][]{\href{http://adsabs.harvard.edu/abs/#3}%
    {\def\hyper@linkstart##1##2{}%
     \let\hyper@linkend\@empty\citet[#1][#2]{#3}}}
  \newcommandtwoopt{\citeyearads}[3][][]%
    {\href{http://adsabs.harvard.edu/abs/#3}
    {\def\hyper@linkstart##1##2{}%
     \let\hyper@linkend\@empty\citeyear[#1][#2]{#3}}}
\makeatother
\usepackage{orcidlink}
\usepackage{subcaption}
\usepackage{longtable}
\usepackage{supertabular}
\usepackage{booktabs}
\usepackage{pdflscape}
\usepackage{xltabular}

\authorrunning{Walker, M. H., et al.}
\begin{document}

   \title{ Radial Dependency of ICME-associated Particle Acceleration Processes: Statistical Multipoint Observations from 2016-2023}

   \author{Malik H. Walker \orcidlink{0000-0003-4380-3519}
          \inst{1}
          \and
          Robert C. Allen \orcidlink{0000-0003-2079-5683} \inst{2}
          \and
          George C. Ho \orcidlink{0000-0003-1093-2066}\inst{2}
          \and
          Glenn M. Mason \orcidlink{0000-0003-2169-9618} \inst{3}
          \and
            Christina M. S. Cohen \orcidlink{0000-0002-0978-8127}
\inst{4}
            \and
            Christina Lee \orcidlink{0000-0002-1604-3326}
 \inst{5}
            \and
            Christian Möstl \orcidlink{0000-0001-6868-4152} \inst{6}
            \and
            Emma E. Davies \orcidlink{0000-0001-9992-8471} \inst{6}
            \and
            Eva Weiler \orcidlink{0009-0004-8761-3789} \inst{6,7}}

   \institute{Johns Hopkins University,
             Baltimore, MD, USA 21218\\
              \email{mwalke85@jhu.edu}
        \and
            Southwest Research Institute, San Antonio, TX, USA 78238
         \and
             Johns Hopkins Applied Physics Laboratory, Laurel, MD, USA 20723-6099
          \and
          California Institute of Technology Pasadena, CA, USA 91125
          \and 
          Space Science Laboratory Berkeley, CA 94720
          \and
          Austrian Space Weather Office, GeoSphere Austria, Graz, Austria,
          \and
          Institute of Physics, University of Graz, Graz, Austria
          }

   \date{Received XXXX; accepted XXXX}

 
  \abstract
  {During the propagation of interplanetary coronal mass ejections (ICMEs), evolution of the ICME-driven shock along with interactions with other solar wind structures, planetary bodies, and general changes to their morphology can alter particle acceleration efficiency and transport effects at their associated shocks. While the underlying mechanisms for these processes have been studied, the connection between the radial evolution of the ICME-driven shock during propagation and resulting gradual Solar Energetic Particle (SEP) and Energetic Storm Particle (ESP) intensities, composition, and acceleration has yet to be fully understood. The current distributed array of spacecraft at varying heliocentric distances provides a welcome opportunity to statistically analyze the radial dependency of particle populations and acceleration mechanisms present at ICME-driven shocks. We compile a database of 39 multipoint ICME events from 2016--2023, which are observed in situ by at least two of the following spacecraft: Parker Solar Probe (PSP), Solar Orbiter, ACE, Wind, and STEREO-A. Using the magnetic field, plasma, and ion compositional data provided by these spacecraft, we derive both local shock and ESP spectral shape parameters. By comparing the changes in these parameters at different stages of ICME propagation, we analyze the connection between the evolution of the local shock conditions and the spectral shape. We find evidence to suggest a consistent increase in shock acceleration efficiency with heliocentric distance while the parent ICME is within 0.7~au, followed by a reduction in shock efficiency at further distances.
}


   \maketitle
%

\section{Introduction}\label{intro}

Coronal mass ejections (CMEs) are key contributors to the energetic particle environment within the heliosphere. As they move into interplanetary space, becoming interplanetary CMEs (ICMEs) in the process, they can drive gradual Solar Energetic Particle (SEP) events through their formation of shocks \citep[e.g.,][]{Reames2021}. These events contain particles with energies ranging from 30 keV to a few GeV, and often last for days. For particularly fast ICMEs, the arrival of these initial SEPs is followed by an Energetic Storm Particle (ESP) event: a large increase in relatively low energy particles (above $\sim$ 0.05 MeV/nucleon) locally accelerated at the driven shock and associated with its crossing. Since large influxes of energetic particles can pose a significant threat to microelectronics on satellites and human spaceflight \citep[e.g.,][]{Feynman2000}, the prediction of the intensity and energy of ESPs are a current priority in the field of space weather \citep[e.g.,][]{tsu2003,garcia2016, giac2020}. Additionally, while SEPs and ESPs are related, sharing the same underlying acceleration mechanism, the effects of transport through the interplanetary medium seen in SEP populations are essentially negligible for ESP events due to their locally accelerated and trapped nature. The negligibility of such effects makes in situ measurements of ESP events especially effective for probing particle acceleration and identifying seed populations of previously accelerated particles \citep[for a review on SEPs and ESPs, see][]{desaiandgiac2016}.

The basic mechanism behind the acceleration of ESPs at its associated shock is generally accepted to be Diffusive Shock Acceleration \citep[DSA; for a review see][]{dru1983}. During this process, particles are trapped within the converging upstream and downstream regions of the plasma surrounding the shock front. Through diffusive effects, the particles are then allowed to scatter between these two boundaries, gaining energy with each interaction. In theory, DSA dictates the shape of the resulting spectra for the accelerated particles, predicting a power law spectra that is solely dependent on the shock compression ratio, or the ratio between the downstream and upstream plasma density. However, observed particle spectra for ESP events often contain points of significant slope change, or spectral breaks, that are more accurately modeled by either the Ellison-Ramaty (ER) function \citep{er1985} or a variation of the double power law \citep{band1993}. While the origin of these spectral breaks remains unclear, with possible causes including finite shock-size, finite acceleration time, particle escape, or transport effects, they have been shown to be systematically dependent on the charge per mass ratio (Q/M), and by extension the diffusion coefficient \citep{li2009}. As spectral breaks divide the spectra into two power law slopes, they are integral for determining the intensity of high energy particles associated with the passage of ICMEs, making them an important feature for SEP prediction models to accurately estimate particle fluences.

Spectral break energies are also expected to change as the ICME-driven shock propagates through interplanetary space. Typically, both the strength of the solar magnetic field and the speed of the ICME shock decreases with increasing distance from the Sun. As such, ICME-driven shocks at further heliocentric distances should have a corresponding increase in both the particle acceleration timescale and the associated diffusion coefficient, which reduces the maximum energy that the shock can accelerate particles to while also allowing for the escape of high energy particles from the shock region. The result is the decrease of the spectral break energy as a function of radial distance \citep{liandzank2005,kong2017}. This effect has been seen in a number of numerical studies. For example, \citet{hu2018} produced simulated particle spectra at different points in the heliosphere for an ICME-driven shock, finding that the decrease in spectral break energy should be the most noticeable in the case of radially aligned (along the same Sun-Earth line) spacecraft. Additionally, \citet{chen2022} simulated particle acceleration along an ICME-driven shock with varying local magnetic field as it propagated to 1~au, also noting a decrease in maximum energy with increasing distance from the Sun. However, until recently this effect had been difficult to definitively observe in situ, as it would require spacecraft capable of conducting robust particle spectrometry placed at varying heliocentric distances to take multipoint observations of ICME events. Although now with the launch of missions like Parker Solar Probe \citep[PSP;][]{pspfox2016} and Solar Orbiter \citep{solOmuller2020} simultaneously probing different regions of the inner heliosphere along with concurrent observations near 1~au, such a study of radial impacts to ICME-driven shock acceleration is more accessible than ever.

\citet{walker2025} presented an in-depth analysis of the 10 March 2022 ICME using in situ measurements from Solar Orbiter \citep{solOmuller2020} and the Advanced Composition Explorer \citep[ACE;][]{acestone1998}, finding evidence of radial effects on the evolution of \textsuperscript{4}He spectra between 0.45 and 1~au. In this study, we extend this analysis to 23 additional multipoint events to provide a statistical analysis of the radial dependence of ICME-associated shock acceleration. This paper is organized as follows. Sections \ref{instruments}  and \ref{select} give an overview of the database, and the methods by which we determined the shape of the collected ESP spectra are given in Section \ref{spectra}. Section \ref{results} presents the results of our analysis of these spectral shape parameters, and Section \ref{discuss} discusses these results. Conclusions are given in Section \ref{conclusion}.

\section{Methodology}\label{methods}

\subsection{Instruments}\label{instruments}
To study the radial evolution of ICME-driven shock associated particle acceleration, this study primarily utilizes energetic particle data taken from Parker Solar Probe, Solar Orbiter, ACE, and the STEREO-A spacecraft. Plasma, magnetic field, and remote sensing observations are also used to connect multipoint observations of each ICME passage.

Parker Solar Probe was launched in 2018 August with the goal of providing in situ measurements of the outermost part of the corona. It has managed to travel within 0.045~au from the Sun. Magnetic field data is taken from the fluxgate magnetometer on the FIELDS instrument suite \citep{pspbale2016}, and bulk plasma data is provided by the Solar Probe Cup \citep[SPC;][]{pspcase2020} onboard the Solar Wind Electrons, Alphas, and Protons \citep[SWEAP;][]{pspkasper2016} suite. Measurements of particle energy and flux are taken from the Energetic Particle Instrument - Low energy (EPI-Lo) onboard the Integrated Science Investigation of the Sun \citep[IS$\odot$IS;][]{pspmccomas2016} suite, which captures ion with energies from $\sim$ 20~keV~nuc\textsuperscript{-1} to 15~MeV~nuc\textsuperscript{-1}. The only species with publicly available flux products from EPI-Lo are H and \textsuperscript{4}He.

Solar Orbiter launched in 2020 February, and has since relayed valuable information regarding the solar wind plasma in the inner heliosphere, reaching distances within 0.3~au. Thermal ion velocity, density, and temperature measurements are taken from the Solar Wind Analyser - Proton-Alpha Sensor \citep[SWA-PAS;][]{soloowen2020}, and magnetic field data is taken from the fluxgate magnetometer \citep[MAG;][]{solohorbury2020}. Energetic particle flux and energy data are taken from the Energetic Particle Detector \citep[EPD;][]{solorodriguez2020, solowimmer2021}, which consists of the SupraThermal Electons and Protons (STEP), Electron Proton Telescope (EPT), Suprathermal Ion Spectrograph (SIS), and High-Energy Telescope (HET) sensors. Ion compositional data is specifically taken from SIS, which covers a broad range of energies (14~keV~nuc\textsuperscript{-1} to 20.5~MeV~nuc\textsuperscript{-1}) and ion species, of which we use H, \textsuperscript{4}He, C, O, and Fe flux products for this study.

The twin spacecraft of the Solar Terrestrial Relation Observatory \citep[STEREO;][]{stakaiser2008} mission launched in 2006 October into orbit at 1~au, with STEREO-A situated ahead of and STEREO-B trailing behind Earth's orbit. While the contact with STEREO-B spacecraft was lost in 2014, STEREO-A currently still provides both in situ and remote sensing observations. As such, only observations from STEREO-A are used in this study. Bulk solar wind measurements are taken from the PLAsma and SupraThermal Ion Composition investigation \citep[PLASTIC;][]{stereogalvin2008} and ion flux data is taken from the Solar Electron Proton Telescope \citep[SEPT;][]{stereomuller2008}. Energetic particle data for the H, \textsuperscript{4}He, C, O, and Fe populations is taken from the Suprathermal Ion Telescope \citep[SIT;][]{stereomason2008} onboard the IMPACT suite \citep{stereoluhmann2008} over an energy range of $\sim$~0.04--9~Mev~nuc\textsuperscript{-1}. Magnetic field data is taken from the fluxgate magnetometer \citep[MAG;][]{stereoacuna2008}. Remote sensing data from the outer coronagraph COR2A and the Heliospheric Imager (HIA) onboard the Sun Earth Connection Coronal and Heliospheric Investigation suite \citep[SECCHI;][]{stereohoward2008} is also utilized in this study.

The Wind \citep{windharten1995} spacecraft was launched in November 1994 to orbit the magnetosphere. This orbit was adjusted in November 1996, placing the spacecraft in halo orbit around L1. The magnetic field data is taken from the Magnetic Field Investigation instrument \citep[MFI;][]{windlepping1995}.  Solar wind thermal ion data is taken from the Solar Wind Experiment \citep[SWE;][]{windogilvie1995}, while energetic ion energy and flux data is taken from the 3D Plasma Analyzer \citep[3DP;][]{windlin1995}.

The ACE spacecraft launched on 1997 August to orbit at L1, offering measurements of solar wind plasma as well as warnings of geomagnetic storms. Energetic particle flux and energy data taken from the Ultra-Low Energy Isotope Spectrometer \citep[ULEIS;][]{acemason1998} are used to supplement the plasma and magnetic field data taken by Wind, allowing for compilation of heavy ion spectra at Earth over a range of energies (50~keV--5~MeV). ULEIS is specifically optimized for species from \textsuperscript{4}He - Fe , exhibiting a low efficiency ($<1\%$) for H. As such we only use data for the \textsuperscript{4}He, C, O, and Fe particle populations.


The Solar and Heliospheric Observatory \citep[SOHO;][]{sohodomingo1995} launched in 1995 December into orbit around L1, with the purpose of conducting remote sensing observations of the near-solar environment. In this study, we use CME kinematics calculated from contextual observations of the corona taken from the Large Angle Spectroscopic Coronagraph \citep[LASCO;][]{sohobrueckner1995}. 

\begin{figure*}[t]
        \centering
        \includegraphics[width = 1\linewidth]{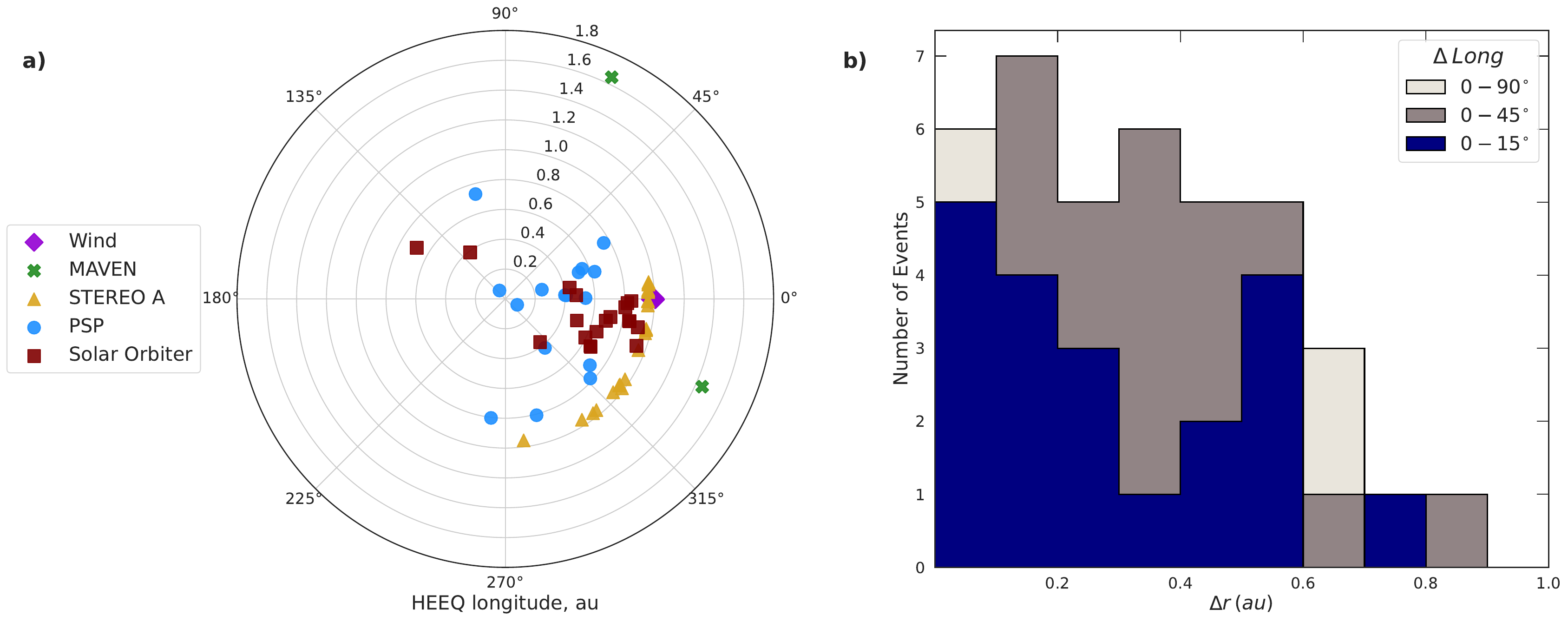}
    
    \caption{Overview of the multipoint event catalog. Panel a): Spacecraft locations in the Heliocentric Earth Equatorial (HEEQ) system for each event. Panel b): Radial and longitudinal separation between observing spacecraft for each event.}
    \label{fig:catalog}
\end{figure*}

\subsection{Event Selection}\label{select}

To identify multipoint observations of ICME events seen by these spacecraft, we start by using the HELIO4CAST ICME catalog, ICMECAT \citep{mostl2020}. The ICMECAT includes in situ observations from the solar wind observatories Wind, STEREO A/B, PSP, Solar Orbiter, BepiColombo, Juno, Venus Express (VEX), MErcury Surface, Space ENvironment, GEochemistry and Ranging (MESSENGER), Ulyssess, and the Mars Atmosphere and Volatile EvolutioN (MAVEN), making for a combined total of 1700 identified ICME observations. We note that while this catalog is periodically updated \citep{mostl2026}, this study only includes entries up to January 2024. With this collection of ICME observations, we were then able to match the individual spacecraft observations to others that were observing the same ICME.

In order to match the events, we identified possible multipoint events within the data provided by ICMECAT. As the typical ICME propagation time from the Sun toward Earth at 1~au is 2-4 days \citep[e.g.,][] {gopalswamy2001, richardsoncane2010, iwai2021}, we grouped subsequent individual identifications whose observation times were less than 3 days apart. We conducted by-eye validation of in situ magnetic field and plasma observations, as well as relative spacecraft positions, to further confirm ICME shock sightings and determine whether a multipoint observation could have been made given spacecraft longitude and ensure the event order was consistent with spacecraft location. Most ICMEs exhibit widths $\leq 120^{\circ}$ , with the majority having widths below $60^\circ$ \citep{gopalswamy2009}. Based on these typical widths, we excluded events with a total longitudinal separation greater than 90 degrees between the observing spacecraft. As the goal of this study is to understand the spatial variability of ESP populations, we also excluded events with a total radial separation of less than 0.03~au.

To verify the preliminary multipoint event classifications, we used a handful of different methods, a number of which were informed by LineupCAT \citep{mostl2022}, a catalog within the HELIO4CAST suite of multipoint observations of ICMEs taken both in situ and remotely. These methods mainly involve the use of contextual simulations of solar wind speed and remote sensing observations.

Solar wind simulations are obtained via the online tool developed at the Community Coordinated Modeling Center (CCMC). This tool utilizes space weather and remote sensing data from the Space Weather Database of Notifications, Knowledge, Information (DONKI) to provide initial conditions for the WSA-ENLIL+Cone model, a 3D magnetohydrodynamic modeling system capable of describing the evolution of the kinematic properties of CMEs in response to the background solar wind plasma and magnetic field. The result is a comprehensive picture of ICME passage, complete with possible arrival times at both planetary bodies and spacecraft with available position data. The online database contains simulation results, and in the case of a possible in situ sighting, simulated arrival times are given with a corresponding margin of error, typically on the order of hours. If a multipoint observation that we identified was made within the margin of error given by the simulation, it was kept as a potential event. In some instances, the online database linked in situ observations to a specific CME event in addition to simulation results, improving this portion of the verification process.

Remote sensing observations are used to link CME eruptions to in situ observations of their interplanetary counterparts through the use of the ARRCAT database, which is compiled as a part of the HELIO4CAST suite \citep{mostl2017}. It contains predicted ICME arrival times at various spacecraft and planetary bodies including Earth-L1, Mercury, Venus, Mars, STEREO-A, STEREO-B, Solar Orbiter, PSP, BepiColombo, VEX, MESSENGER, Ulysses, Juno, and JUICE. These arrival times are computed through the addition of the CME eruption time and a travel time based on CME kinematics sourced from HIGeoCAT \citep{barnes2019}, a CME catalog that derives these kinematics through the use of single spacecraft geometric fitting models applied on observations from STEREO/HI. The specific model results used to compile ARRCAT are created using the Self-Similar Expansion model \citep[SSE; see][for a review]{davies2013}. For verification, the model results added confidence that flagged events originated from the same CME, with multipoint sightings confirmed when in situ observations of the event occurred within 24~hr of their respective modeled arrival times.

In cases where an individual in situ observation could not be linked to a multipoint event using the ARRCAT catalog, or where all observations of a multipoint event are missing from said catalog due to the initial CME eruption being missed by HIA, we use the DONKI database as a supplement. In addition to data from solar wind simulations, CME database entries in the DONKI database also have information derived from remote sensing observations, such as the source location or a contextual description of the plasma environment. In some cases, these entries are linked to observations of interplanetary shocks detected at Earth. If these shocks are also present in the preliminary multipoint catalog, then those individual observations are considered verified. Otherwise, we input the kinematic properties of the CME derived from remote sensing observations from either STEREO/COR2 or SOHO/LASCO into the Solar-MACH tool \citep{gieseler2023} which is a Python-based application that uses the Parker magnetic field lines in conjunction with ephemeris data from \href{https://ssd.jpl.nasa.gov/horizons/}{JPL Horizons} to visualize the position and solar magnetic connectivity of various spacecraft and planetary bodies within the heliosphere. Using this tool, we can confirm whether a multipoint observation would be plausible based on spacecraft positions and the direction of CME propagation. 

Finally, we cross reference our multipoint events with those found in LineupCAT to provide further validation and additional information for identical entries if applicable. As a result, we compile a catalog of 39 multipoint events seen by a subset of spacecraft from January 1 2016 to December 30 2023 (Table~\ref{tab:fullcatalog}). Figure \ref{fig:catalog} provides an overview of these events, with the left panel illustrating the locations at which each spacecraft included in the catalog observed an event in Heliocentric Earth Equatorial (HEEQ) coordinates. The right panel displays a histogram of the maximum longitudinal ($\Delta Long$) and radial ($\Delta r$) separation between the observing spacecraft for each event in the catalog (i.e., the largest spatial span of the constellation of all observing spacecraft for a given event). Since the focus of this study is on the analysis of the radial evolution of shock associated acceleration, we only use events where the maximum separation spanned by all spacecraft observing a particular ICME was $\Delta r > 0.2$ and $\Delta Long < 45^{\circ}$. We also note that while there are two MAVEN sightings included in the reduced multipoint catalog, due to the lack of energetic particle data for these specific events, we do not include them in the analysis. From these alterations, the number of events analyzed in this study is reduced to 23.

\subsection{ESP Spectral Fitting}\label{spectra}

\begin{figure*}
    \centering
    \includegraphics[width=0.85\linewidth]{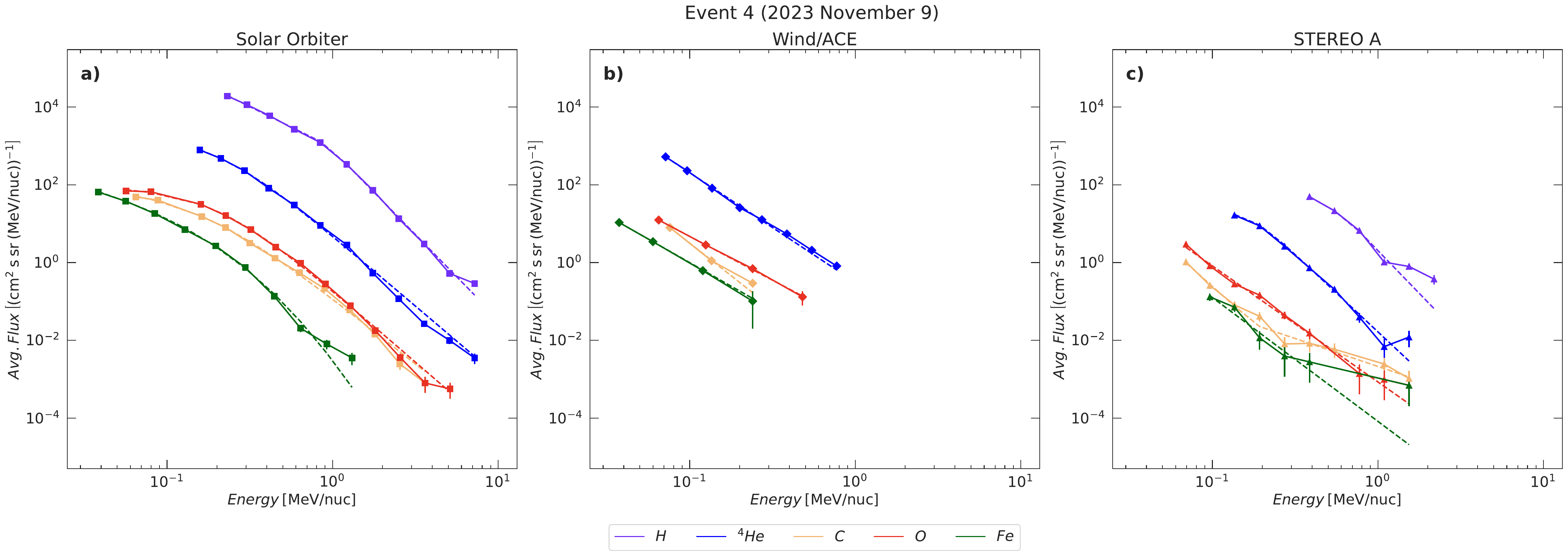}
    \caption{Particle spectra for the passage of ESPs associated with the 2023 November 9 CME. This event was seen Solar Orbiter at 0.69~au, STEREO-A at 0.97~au, and Wind/ACE at 0.98~au. There was a total longitudinal separation of 18.36$^\circ$ between the spacecraft. Panels a) through c) show the spectra observed by Solar Orbiter, Wind/ACE, and STEREO-A respectively. The solid lines denote the observed spectra, while the dotted lines show the fits for these spectra using one of the three models described in Section \ref{spectra}}
    \label{fig:examplecase}
\end{figure*}

ESPs, the population of particles local to the shock structure, allow a direct analysis of the underlying shock acceleration processes and are often consistent with the DSA theory \citep{lario2003,desai2004}. These particles are commonly divided into an upstream and downstream region relative to the shock during an ESP event. Downstream ESPs typically exhibit spectral shapes that differ from those of their upstream counterparts as they are influenced by different processes. In particular, particle escape tends to occur upstream of the associated shock, whereas in the downstream region, magnetic turbulence is more significant. \citet{walker2025} also showed that when comparing the slopes of ESP particle spectra for different ion species, the influence of seed populations, which can alter the ion composition of the event, can be more readily apparent in the upstream ESP population. Thus, for our analysis of the spectral shape of ESP particles, we focus on the upstream portion of the event.

The interval of upstream ESP passage was determined through first identifying intervals dominated by ESPs, defined as the time range of enhanced particle flux surrounding the shock. The ICME-driven shock was located through bulk plasma measurements, namely speed, density, temperature, and ion flux, as well as magnetic field observations. The portion of the ESP dominant interval prior to shock passage was then considered as the upstream ESP population. Average upstream ESP spectra for multiple ion species, namely H, \textsuperscript{4}He, C, O, and Fe were computed (see example in Figure~\ref{fig:examplecase}). These averages were only computed for energy bins within the spectrograms of observing spacecraft that had reliable counting statistics. As discussed in Section \ref{instruments}, H fluxes are excluded for ACE/ULEIS, and the only publicly available species for PSP/EPI-Lo at the time of this study were H and \textsuperscript{4} He. 

To derive the particle spectra shape parameters, we utilize LMFIT \citep{lmfit2025}, a Python package specialized for non-linear least-squares curve fitting and minimization that employs the Levenberg-Marquardt as its default fitting method. This study uses three possible spectra fitting equations to determine the best representation of a given observation. 

The first is the Pan-Spectrum Fitting formula derived by  \citet{liu2020}: \begin{equation}
J(E)=A \times E^{-\beta_1}\left[1+\left(\frac{E}{E_0}\right)^\alpha\right]^{\frac{\beta_1-\beta_2}{\alpha}},
\end{equation}
where $J$ is the particle flux, $E$ is the particle energy, $A$ is the amplitude coefficient, $E_0$ is the spectral break energy, $\alpha$ represents the sharpness and width of the spectral transition region, which is centered around $E_0$, and $\beta_1$ ($\beta_2$) are power-law indices that describe the spectral slope before (after) the spectral transition region. As the Pan-Spectra fitting formula incorporates a number of functions typically used to fit different types of energetic particle spectra as special cases, it is generally applicable for suprathermal particle populations. However, due to the relatively large number of parameters in this formula, and the fact that only a portion of the total particle spectra is observed within the energy range of a given instrument, there are cases in which the Pan-Spectrum Fitting Formula is not sufficiently constrained and therefore not always as accurate as other models. 

The second fitting formula is the classic double-power law (CDPL), \begin{equation}J(E)= \begin{cases}A \times\left(\frac{E}{E_0}\right)^{-\beta_1}, & E<E_0 \\ A \times\left(\frac{E}{E_0}\right)^{-\beta_2}, & E>E_0\end{cases}\end{equation}
using the same variable notation as the Pan-Spectra fitting formula. Although the CDPL does not include a transition region, but rather has a sharp spectral break at $E_0$, for some events it can better capture the profile, especially when the fit is not well constrained. The CDPL has been long used for fitting solar wind suprathermal particle spectra \citep[e.g.,][]{krucker2009,dayeh2017,Henderson2025}.

Lastly, the third fitting formula used is the single-power law function, \begin{equation}J(E)=A \times E^{-\beta},\end{equation}
where $\beta$ is the power law index. Although the simplest of the fitting functions employed in this study, the single-power law best represents some events, especially ones in which the break energy is not observed within the energy range of the instrument. Similarly to the CDPL, the single-power law has long been used to characterize suprathermal particle spectra \citep[e.g.,][]{krucker2009,dayeh2017, Allen2021}. 

After fitting an upstream ESP spectra using all three models, we compute the reduced chi-square statistic \citep[for a review, see][]{bevington1969}. Since this parameter decreases as the accuracy of the fit increases, the model that provides a spectral fit with the lowest reduced chi-square value is chosen as the most accurate \citep{west2012}. If the Pan-Spectrum Fitting Formula is determined to be the most accurate model, we perform an additional fit using an implementation of the affine-invariant ensemble sampler for the Markov Chain Monte Carlo (MCMC) described in \citet{emcee2013}. This is done because the complexity of this specific model typically requires a more robust exploration of the posterior distribution to better characterize parameter uncertainties. Through this implementation we are able to more confidently define model parameters despite the restrictive energy range of the spectrometers onboard the observing spacecraft, resulting in increasingly accurate fits.

Figure \ref{fig:examplecase} shows an example of the results of the fitting process for a three-point observation of an ICME. This event was initially observed remotely on 2023 November 09 by STEREO-A/SECCHI, and was later detected in-situ by Solar Orbiter, STEREO-A, and at L1 by both Wind and ACE. The solid lines in each panel illustrate the observed upstream ESP particle spectra, while the dotted lines show the fitted spectral form for a given spacecraft and species using the model that has the lowest chi-square statistic, as discussed above. We note that there is generally good agreement between the observations and the best fit models.

\begin{figure}
    \centering
    \includegraphics[width=1\linewidth, height = 6cm]{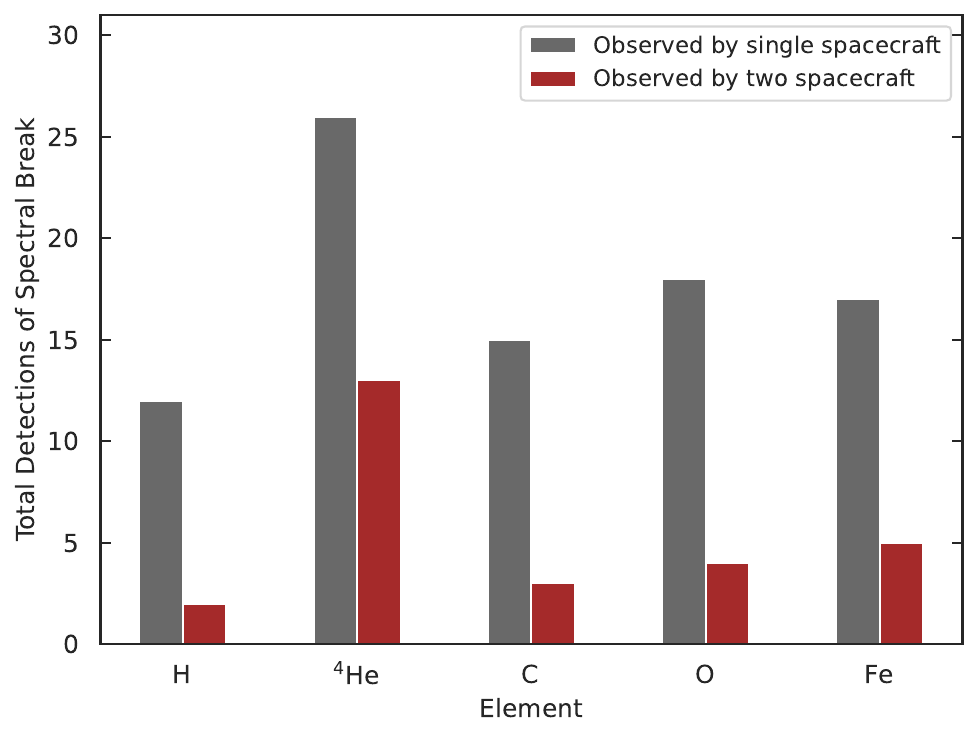}
    \caption{Summary of the total single point and two-point observations of break energy in fitted particle spectra per ion species.}
    \label{fig:breakcount}
\end{figure}

From these methods, we are able to derive the spectral shape parameters for the 23 multipoint ICME events with favorable radial and longitudinal separations to facilitate the analysis of the behavior of the spectral break energy during ICME propagation (equating to 52 individual sightings). A selection of shape and shock parameters for these events are shown in Table \ref{tab:parameters}. Across all single point observations of these events, 17 \textsuperscript{4}He spectra were best fit by the Pan-Spectrum fitting formula, 9 by the CDPL, 13 by a single power law, and 13 were not well fit. Figure \ref{fig:breakcount} shows the number of events where spectral break energy was observed for a specific ion species, with three-point events being split into 2 two-point observations. From the counts presented, we find that more two point observations of break energy, which correspond to a measurement of break energy evolution for an event, are made for \textsuperscript{4}He than any other species. Therefore, for single point measurement analysis, we can include all species, but for multipoint analysis we only use \textsuperscript{4}He.

\section{Results}\label{results}

\begin{figure*}[h!]
   \centering
            \includegraphics[width=1\linewidth]{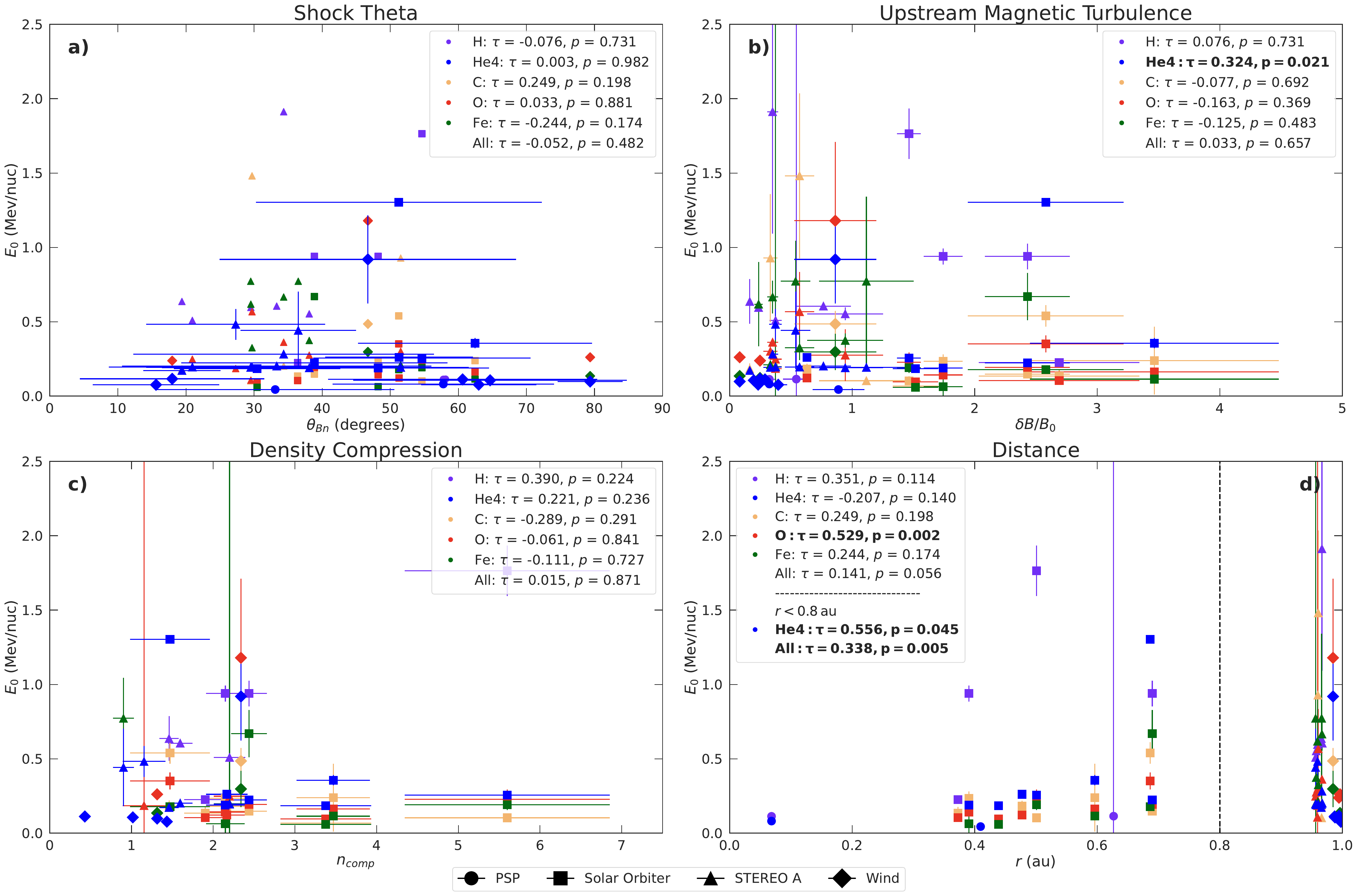}
   \caption{Scatter plots showing the relationship between break energy, $E_0$, and various parameters for individual ICME observations of the 23 multipoint events used for this study. (a) Shock normal angle ($\theta_{bn}$), (b) Upstream magnetic turbulence ($\delta B/B_0$), (c) Density compression ($n_{comp}$), (d) heliocentric distance ($r$). For each data point, observing spacecraft is designated by shape, and particle species is designated by color. The calculated Kendell's tau coefficient ($\tau$) and p-values ($p$) are shown in each panel, with statistically significant relationships shown in bold. For clarity, uncertainties for $\theta_{Bn}$ and break energy in panel a are only shown for \textsuperscript{4}He, with a more robust depiction provided in Appendix~\ref{app:thetaunc}.}
              \label{fig:kendelltau}%
    \end{figure*}

Figure~\ref{fig:kendelltau} shows the assessment of any possible correlation between the break energies (for spectra that are best fit by either the Pan-Spectral fitting formula or CDPL) and local shock parameters, namely shock theta ($\theta_{bn}$, panel a), upstream magnetic turbulence ($\frac{\delta B}{B_{0}}$, panel b), and density compression ($n_{comp}$, panel c), as well as heliocentric distance (panel d). The calculation of these parameters are discussed in Appendix \ref{app:shockcalc}, and the uncertainties for non-\textsuperscript{4}He data points in Figure~\ref{fig:kendelltau}a are shown in Appendix \ref{app:thetaunc}. To quantify how well two values are correlated we use the Kendall’s $\tau$ coefficient, which is a non-parametric measure of ordinal association that is well suited for determining relationships between variables in small datasets due to its decreased sensitivity to outliers and skewed distributions \citep{kruparova2025,kendall1938}. We also calculate the corresponding $p$ values to help classify correlations that are statistically significant, which we define as cases where $|\tau| > 0.25$ and $p<0.05$. Correlations that satisfy this condition are shown in bold.

From Figure~\ref{fig:kendelltau}a we find that there are no species in which the correlations between the spectral break energy and a local ($\theta_{bn}$) are considered to be significant, suggesting that local ($\theta_{bn}$) is not the sole determinator for a spectral break energy. Similarly, Figure~\ref{fig:kendelltau}b shows that most species do not see a systematic correlation between the break energy and the upstream magnetic turbulence, with the exception of \textsuperscript{4}He. Figure~\ref{fig:kendelltau}c again depicts a lack of significant correlation between break energy and the shock density compression across all species. However, it is noted that there is more scatter in break energy for lower ($<3$) compression ratios, with high compression ratios ($>3$) having break energies below $\sim0.4$~MeV/nuc. Figure~\ref{fig:kendelltau}d shows that oxygen had a significant correlation between break energy and the heliospheric distance of the observation, though this could be due to a non-monotonic relationship between these values. In fact, when the $\tau$ coefficients shown in Figure~\ref{fig:kendelltau}d were recalculated after removing events that were observed at distances greater than 0.8~au, the correlation across all ion species became statistically significant, with $\tau = 0.338$ and $p = 0.005$. The correlation between distance and break energy for \textsuperscript{4}He measurements, specifically, was also significant after these recalculations, with $\tau = 0.556$ and $p = 0.045$. 
Thus, despite the large event-to-event variability, the maximum break energy seems to increase from near-Sun observations out to $\sim$0.7~au. Together, these results suggest that the use of any one local shock parameter to estimate the spectral break energy is insufficient. 

\begin{figure*}[h!]
    \centering
    \includegraphics[width = 0.85\linewidth]{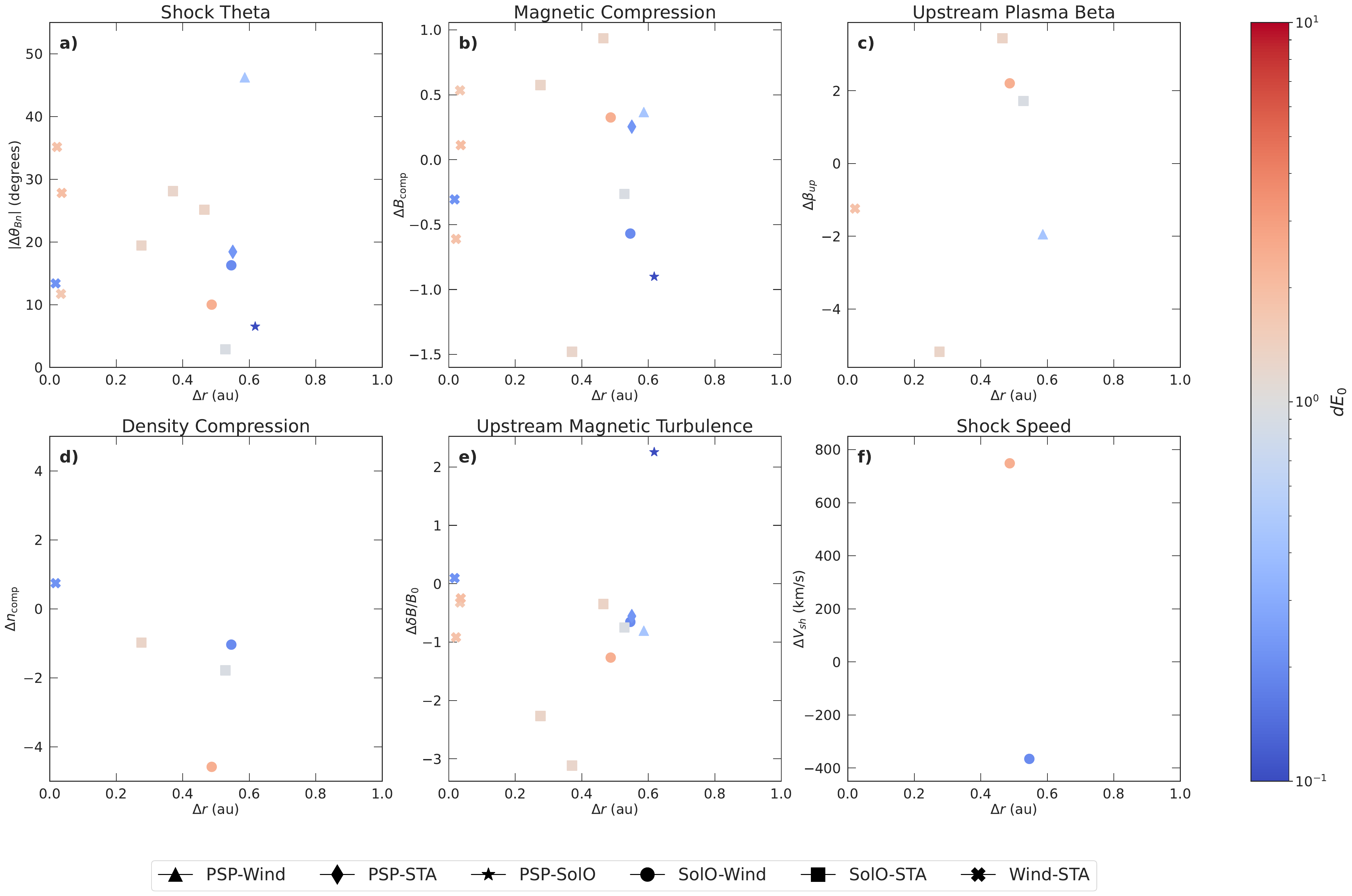}
    \caption{Scatter plots showing the correlation between changes in heliocentric distance ($r$), various shock parameters, and spectral break energy ($E_0$) for two-point \textsuperscript{4}He observations of upstream ESPs associated with ICME events. (a) Shock normal angle, (b) magnetic compression ($B_{comp}$), (c) upstream plasma beta ($\beta_{up}$), (d) density compression, (e) upstream magnetic turbulence, (f) shock speed ($V_{sh}$). The change in spectral break energy ($dE_0$) is defined using Eq. \ref{eq:breakchange}. Red (blue) points indicate a decrease (increase) in break energy with distance.}
    \label{fig:2pointpair}
\end{figure*}
\begin{figure*}[h]
    \centering
    \includegraphics[width = 0.85\linewidth]{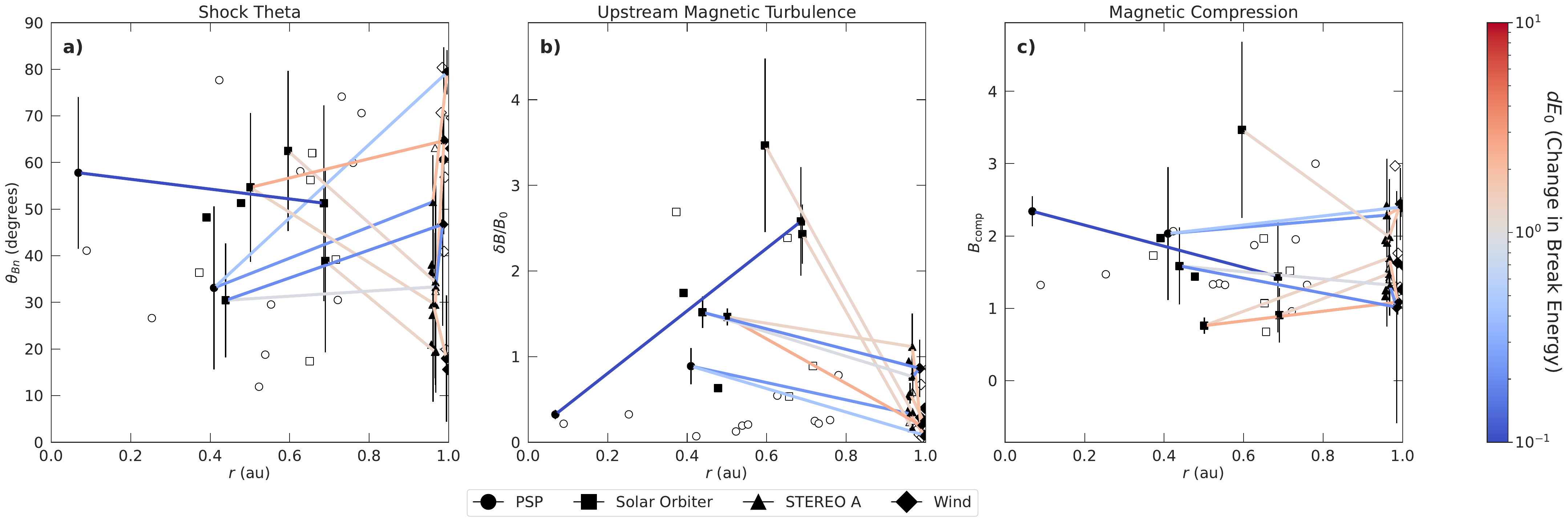}
    \caption{Plots showing the correlation between heliocentric distance ($r$), shock parameters, and break energy ($E_0$) for individual \textsuperscript{4}He measurements of upstream ESPs associated with multipoint ICME events. (a) Shock normal angle ($\theta_{bn}$), (b) upstream magnetic turbulence ($\delta B/B_0$) , and (c) magnetic compression ($B_{\mathrm{comp}}$). For each point, the observing spacecraft are designated via shape. The colored lines connect observations of the same ICME event, and the color is determined by the value $dE_0$ defined in Eq. \ref{eq:breakchange}. Similar to Figure \ref{fig:2pointpair}, red (blue) points indicate a decrease (increase) in break energy with distance. Points without colored lines denote events where only one spacecraft observed a spectral break, while hollow points denote cases where the spectral breaks were not observed.}
    \label{fig:coloredlines}
\end{figure*}

We attempt to remove some of the event-to-event variability by investigating the relative change in parameters for conjunction events of the same ICME. Through comparing changes between observations of the same ICME, any systematic difference between the events may become effectively normalized out. As such, Figure~\ref{fig:2pointpair} shows the relative change in spectral break energy (indicated by color) with respect to the change in heliocentric distance and the variations in the local shock parameters between each identified two-point pairing of observations from the multipoint events catalog (a list of parameters for these events are given in Table \ref{tab:parameters}). This analysis was done only for the \textsuperscript{4}He spectra due to the relatively large number of instances of spectral break observation, as discussed in Section \ref{spectra}. The change in spectral break energy is defined as: \begin{equation}
    dE_0 = \frac{E_{0,1}}{E_{0,2}}, \label{eq:breakchange}
\end{equation} where $E_{0,1}$ and $E_{0,2}$ are the break energies seen by the observer closer to and further away from the Sun, respectively, in a two-point measurement pair. This means that events where the break energy decreases with distance are red, while instances where the opposite occurs fall along the blue end of the spectrum. We note that the lack of data points in some panels is due to missing plasma data from the observing spacecraft, as parameters such as $V_{sh}$ require plasma velocity and temperature, and calculating the change in this value between observers requires this data to be available at both. Three-point events are also split into two pairs of two-point events for this analysis. 

An advantage of displaying the data in this way is that points along a horizontal line indicate variations due to radial evolution at a fixed difference in local parameter, while vertical variations indicate changes due to changed in the local parameter for a fixed radial difference in observations. As such, according to the prevailing theory regarding shock evolution within the heliosphere, all the points in Figure \ref{fig:2pointpair} along the horizontal 0-line (i.e., no change in the local shock parameter) should be red, with hue increasing in intensity as the distance between the spacecraft ($\Delta r$) increases, indicating a gradual lowering of the break energy with distance \citep[e.g.,][]{hu2018}. The presence of blue points, specifically at high $\Delta r$, suggests an increase in shock acceleration efficiency between a region near the Sun and 1~au, which should only occur if the local shock parameters change enough, and in the right direction, to increase the acceleration efficiency by more than the expected break energy relaxation with increasing heliospheric distance. 

Panels a-f of Figure \ref{fig:2pointpair}, which depict the variability of $\theta_{bn}$, magnetic compression ($B_{\mathrm{comp}}$), upstream plasma beta ($\beta_{up}$), density compression, upstream magnetic turbulence, and shock speed ($V_{sh}$), respectively, all contain blue points predominantly at high $\Delta r$, with the number increasing with distance. There is also evidence to suggest that this trend occurs regardless of the variation in local parameter. This can be seen most clearly in panels a and b, as for both quasi-parallel and quasi-perpendicular shocks, as well as shocks with positive and negative changes in magnetic compression, increases in acceleration efficiency are seen when $\Delta r \gtrsim 0.5 \: \mathrm{au}$. Panel e also contains a high density of blue points at large spacecraft separations. However, the change in upstream magnetic turbulence is within the range  $-1\leq\Delta \frac{\delta B}{B_{0}}\leq 0$ for most two-point observations, and points outside of this range are at opposite extremes (both in local parameter variation and hue) for relatively low ($\Delta r < 0.5$) and high ($\Delta r > 0.5$) radial separations, making it difficult to determine the effects of variations in magnetic turbulence on shock acceleration efficiency without additional events. Similarly, more events with robust plasma data are required to make definitive conclusions regarding the relationship between local shock parameter variation and acceleration efficiency for panels c, d, and f.

One consideration of the representation of Figure \ref{fig:2pointpair} is that it can only easily allow for the elucidation of variations that are due to monotonic dependence. This means that if a parameter increases the break energy to a point before lowering it, or if the dependence plateaus, then scatter would be injected into this representation. To further probe this possible effect, Figure~\ref{fig:coloredlines} represents the same two-point comparisons as a function of the observed shock parameters versus the heliospheric distance at each spacecraft and the change in break energy being shown as a colored line between these points. Panel a depicts the variation in $\theta_{bn}$, panel b shows the variation in the upstream magnetic turbulence, and panel c the variation in magnetic compression. We only include measurements of this subset of parameters as they were the only ones that could be reliably calculated at a majority of the observing spacecraft as seen by the number of points in Figure~\ref{fig:2pointpair}. Points without colored lines denote events where only one spacecraft observed a spectral break, resulting in an inability to track the evolution of this break energy. Hollow points denote cases where spectral breaks were not observed by a spacecraft—either due to the spectra being best fit by a power law, or due to insufficient particle flux data, which precluded meaningful spectral fitting. We find that instances of break energy increase (blue lines) primarily occur when the first observer is within 0.5~au. This result, along with the positive correlation between break energy and distance shown in Figure~\ref{fig:kendelltau}d, implies that the break energy increases from the Sun to $\sim0.7$~au before beginning to decrease to lower energies as expected. Additionally, the observed changes in break energy do not seem to be primarily dependent on the local shock parameters, as there is no visual correlation between instances of break energy increase or decrease with changes in these parameters (i.e., both blue and red lines exist for increasing/decreasing values of $\theta_{bn}$, upstream magnetic turbulence, and magnetic compression, suggesting the absence of a systematic variation with these parameter).

\section{Discussion}\label{discuss}

\begin{figure*}[t]
    \centering
    \includegraphics[width = 0.85\linewidth]{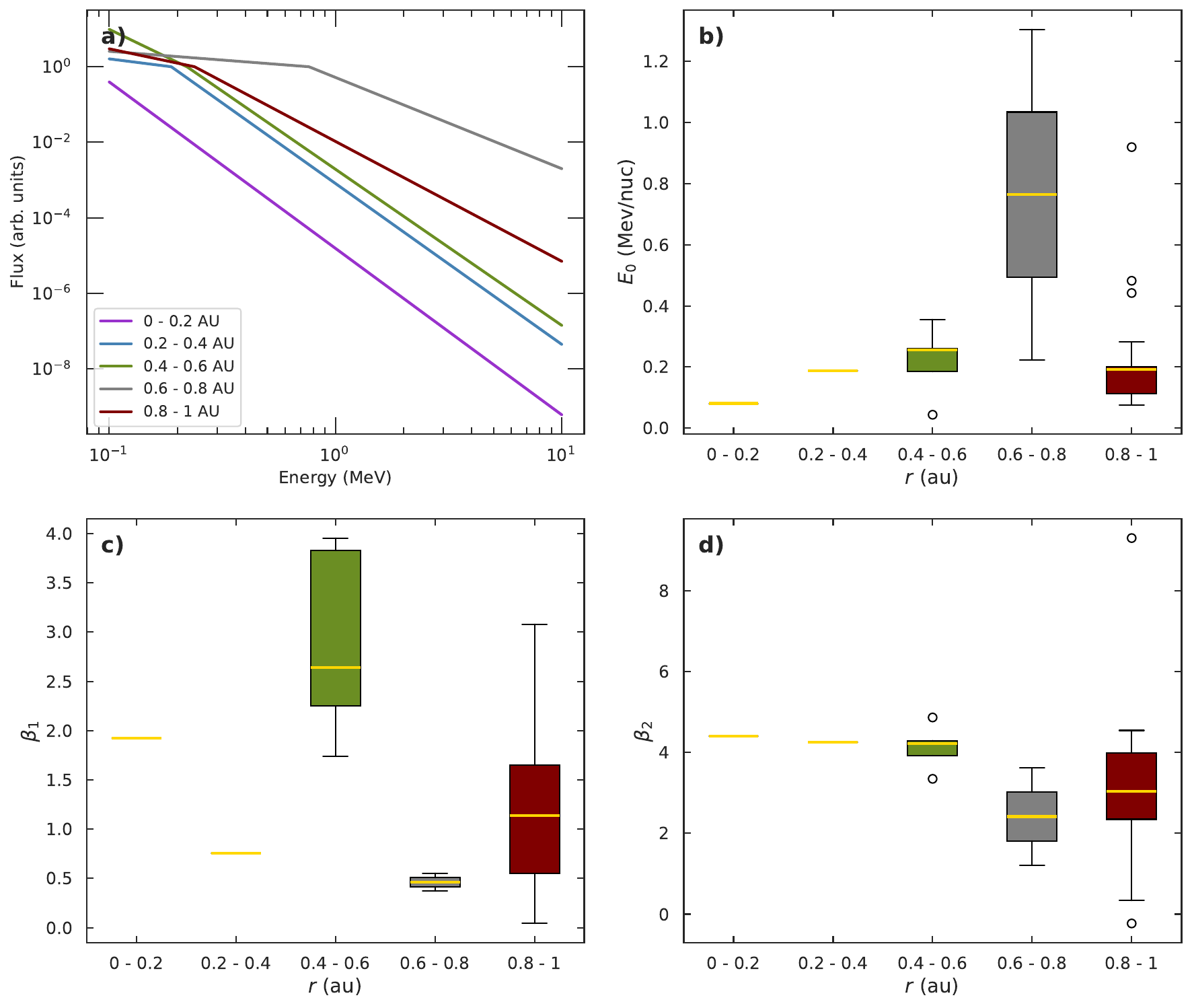}
    
    \caption{Evolution of spectral shape parameters (namely the break energy, and slopes both before and after the break) calculated for upstream ESP, \textsuperscript{4}He measurements and averaged over equally spaced distance bins. (a) Changes in average double power law spectra with distance, (b) - (d) box plots showing the change in break energy, $\beta_{1}$, and $\beta_{2}$, respectively, with distance. The counts in each bin, in order, are 1, 1, 5, 2, and 17. }
    \label{fig:discussion}
\end{figure*}

Figure \ref{fig:discussion} provides a more comprehensive view of the radial evolution of the upstream ESP spectral form. By taking the average fitting parameters for \textsuperscript{4}He particles in the upstream ESP time interval from individual observations within defined radial bins (the midpoints of which are shown in panel a)), we calculate average double power law spectra at different heliocentric distances. For spectra fit by the Pan Spectrum fitting formula, we remove the $\alpha$ parameter from consideration for ease of comparison to the other models. Figure~\ref{fig:discussion}a shows the resulting average spectra, with flux normalized to arbitrary units for ease of comparison. Additionally, Figure~\ref{fig:discussion} panels b, c, and d illustrate the radial evolution of the 25th, 50th, and 75th percentiles for the spectral break energy, $\beta_1$, and $\beta_2$, respectively. Outliers are defined as values exceeding 1.5 times the interquartile range.  A direct relationship between break energy and distance throughout the inner heliosphere is seen (Figure~\ref{fig:discussion}a and b), in that the break energy increases from the Sun to a peak between 0.6--0.8~au before beginning to relax to lower energies further out, decreasing significantly at 1~au. While we note that the innermost two bins are limited statistically with only one event in each, this general trend is also evident in the changes of the inner quartile range for the other three bins. The relaxation can be readily explained as a byproduct of weakening shock strength and magnetic field intensity further from the Sun as discussed in Section~\ref{intro}. However, this theory cannot explain the initial increase in break energy, which suggests a corresponding increase in particle acceleration efficiency in the inner heliosphere. Additionally, while the lower energy spectral index, $\beta_1$, does not have a clear statistical variation by radial distance (Figure~\ref{fig:discussion}a and c), the higher energy spectral index, $\beta_2$, though exhibiting greater variability and a locally elevated median around 1~au, systematically decreases with distance (Figure~\ref{fig:discussion}a and d).

A possible explanation for this observed trend in break energies with heliocentric distance is particle trapping. During transit, ICME-driven shocks induce upstream turbulence, which creates a foreshock region. This region can trap particles near the shock wave, allowing for continuous acceleration over a longer timescale \citep{wijsen2022}. As such, it could be that within the inner heliosphere, particle trapping by the foreshock is efficient enough to increasingly confine high energy particles close to the shock at further heliocentric distances. This would increase the particle intensity at higher energies as these particles accelerate during transit, resulting in an increase in spectral break energy. Such a process would continue until the shock strength and magnetic field intensity decline enough to decrease the efficiency of particle trapping at the foreshock, allowing for a decrease in break energy due to particle escape.

Recent studies have also suggested that the maximum energy an ICME-driven shock is able to accelerate particles up to can vary within the inner heliosphere beyond a simple decrease as the ICME slows and magnetic field weakens. For example, several studies have investigated SEP events with a characteristic delay to their maximum energy, sometimes called Delayed Maximum Energy (DME) events or Inverse Velocity Arrival (IVA) events. These DME/IVA events have been explained through invoking a combination of varying magnetic connectivity \citep[e.g.,][]{Ding2025,allen2026} and/or through changes in the acceleration rate at the shock \citep[e.g.,][]{Cohen2024,chen2025,Xu2026}. For instance, \citet{allen2026} conducted a statistical study of 31 SEP events -- 11 of which were DME/IVA events -- and reported that the occurrence rate of DME/IVA events was larger when the footpoints of Solar Orbiter were westward of the associated flare location, consistent with simulation results of the impacts from changes in magnetic connectivity to an expanding ICME-driven shock \citep{Ding2025}. Further, the location of maximum energy released into the observed flux tube for DME/IVA events could be as large as $\sim0.5$~au, leading to their conclusion that DME/IVA events are a result of both an observer’s evolving magnetic connectivity as well as time-varying particle acceleration as the shock expands into a structured solar wind \citep{allen2026}. 

\citet{chen2025} simulated observations of the 2022 September 5 SEP event seen by PSP at 0.07~au and Solar Orbiter at 0.71~au to investigate potential impacts of time-dependent diffusive shock acceleration. They posited that the cause of the delayed arrival of higher energy particles at PSP was due to temporal and spatial variations in the acceleration rate. Specifically, high upstream solar wind speeds suppressed shock strength at the flank, which is the region that PSP was initially connected to. This increased the time necessary to accelerate particles to higher energies close to the Sun, leading to the delayed production and release of the maximum energy particles, resulting in the DME/IVA signature at PSP. 

A similar near-Sun suppression of shock strength was also reported in \citet{chiappetta2026} for the 2023 March 13 SEP event observed in situ by PSP, Solar Orbiter, and Wind. By combining measurements from these spacecraft, in addition to remote sensing observations taken by STEREO-A/COR2 and SOHO/LASCO/C3, they were able to derive the radial evolution of the density compression ratio and Alfvén Mach number. They find that $M_{A}$ is relatively low near the corona ($\sim~1.2$), but begins to increase by the time the shock crosses PSP at 0.23~au, stabilizing at a value of 4.68 at 0.6~au as seen by Solar Orbiter, before decreasing by almost half near Earth. A similar effect is seen for the density compression ratio, though there is less change between the coronal values and the observed ratio at PSP, likely due to its connection to the shock flank for this event. In each of these studies, higher energy particles are produced in the inner heliosphere rather than in the corona, mainly due to interactions with the upstream plasma.

We note that shock geometry could also contribute to the observed increase in break energy, as certain regions of the shock tend to accelerate particles more efficiently \citep[e.g;][]{chen2022,Ding2025}. However, Figure~\ref{fig:coloredlines}a shows that there is no significant ordering of the break energy change with shock obliquity. Instead, the break energy mostly decreases when the initial observer is past 0.5~au, and the opposite when the initial observer is within 0.5~au, regardless of the obliquities seen at either observer in a pair. This result strengthens the notion that $\theta_{Bn}$ is not a dominant parameter in determining the evolution of the break energy. Additionally, \citet{wijsen2022} simulated low energy (50~keV--2~MeV) particle intensities for a strong ESP event observed by ACE and SOHO at Earth on 2012 July 14. They found that the particle energy spectrum taken at the shock nose for an observer on the Sun-Earth line hardens with radial distance up to $\sim$0.5~au, increasing the break energy in the process. After this point, the spectral shape remains constant as the shock propagates to 1~au. They report that a similar effect is also seen for a virtual observer 30~degrees west in longitude from Earth, suggesting that this process is consistent along a wider area of the shock. The simulated spectra in \citet{wijsen2022} and the average spectra shown in Figure \ref{fig:discussion}a exhibit analogous evolutions of the spectral break energy with distance, both depicting an increase in this energy until around 0.6~au. As the simulations assumed a seed population of suprathermal protons with energies of 50~keV, it is likely that such populations play a key part in the observed spectral behavior. 

It is also important to note that the break energies observed in this study are lower than those found in other works, \citep[e.g;][]{mewaldt2005}, which reported break energies within the energy range \~3 to \~30 MeV/nuc for the five Halloween 2003 ICME events. The difference in break energy range could be due to the choice of time interval over which the spectra were averaged, driven by the aims of particular studies. For example, \citet{mewaldt2005} analyzed particle fluences integrated over the entire ESP event, while this study focuses on upstream ESPs. As mentioned in Section ~\ref{spectra}, the regions up- and downstream of ICME-driven shocks are affected by different processes that can result in noticeable variations in their integrated particle spectra. For example, \citet{Ding2024} compiled two-hour time interval proton spectra for the 30 August 2022 and 05 September 2022 ESP events. They found that the observed spectra changed over the course of each event, with the 05 September event in particular showing a complete reversal in the sign of the spectral index (from positive in the upstream region to negative in the downstream region). Additionally, the particles immediately downstream of the shock generally exhibited higher intensities. By averaging over both shock regions, these variations in spectral shape are minimized, and as such the features in event-integrated spectra may not be representative for upstream ESPs. 

Alternatively, the spectral breaks for events in this study whose spectra are best fit by a power law may lie outside of the measurable energy range of the instruments used. Thus the use of instruments capable of observing higher energy particles could result in more direct observations of spectral break change with radial distance. However, it is unclear if these higher energy breaks would also increase with distance out to 0.7 au.

\section{Conclusion}\label{conclusion}
In this work, we conducted a statistical study of multipoint ICME events to probe the radial effects of shock acceleration. We constructed a catalog of 39 events observed by PSP, Solar Orbiter, Wind, and STEREO-A, focusing our analysis on 23 events that had optimal radial and longitudinal separation between the observing spacecraft. By constructing upstream ESP ion spectra and determining their shape parameters, as well as calculating local shock parameters for each observation, we were able to assess the correlation between changes in these parameters with heliocentric distance. In doing so,  we find that within $\sim0.7$~au the spectral break energy increases with distance, suggesting an increase in acceleration efficiency for shocks in the inner heliosphere. The break energy then decreases as shocks propagate towards 1~au, reflecting the decrease in magnetic field strength and shock speed at these further distances.

We posit that the increase in shock acceleration efficiency close to the Sun is a byproduct of particle trapping within the turbulent foreshock region. This effect would result in the continuous acceleration of high energy particles at the shock during ICME transit through the inner heliosphere. Additionally, the increase in ICME-driven shock acceleration efficiency within 0.7~au may have implications for the interpretation of the delayed arrival of high energy particles characteristic of DME/IVA-type events. Further investigation using a larger number of multipoint ICME events that include in situ observations within 0.7~au is necessary to explore the processes leading to this trend. We also note that while turbulence plays a role in the particle trapping process, we do not observe a significant correlation between the spectral break energy and upstream magnetic turbulence (Figure~\ref{fig:kendelltau}b). However, the results from \citet{Ding2024} suggest that the effects of upstream turbulence are more readily seen in the behavior of the spectral slope, and as such future work will focus on more robustly exploring this relationship. Regardless, this analysis highlights the influence of radial effects on particle energy and intensity that should be taken into consideration in future studies.

\begin{acknowledgements}
Solar Orbiter is a mission of international cooperation between ESA and NASA, operated by ESA. The Suprathermal Ion Spectrograph (SIS) is a European facility instrument funded by ESA during construction, and NASA funds the instrument during Phase E operation. Solar Orbiter post-launch work at SwRI is supported by NASA contract 80GSFC25CA035, and we thank NASA headquarters and the NASA/GSFC Solar Orbiter project office for their continuing support. Parker Solar Probe was designed, built, and is now operated by the Johns Hopkins Applied Physics Laboratory as part of NASA's Living with a Star (LWS) program (contract NNN06AA01C). Support from the LWS management and technical team has played a critical role in the success of the Parker Solar Probe mission. We thank the scientists and engineers whose technical contributions prelaunch have made the IS$\odot$IS instruments such a success. MHW also acknowledges support from the Department of Defense (DoD) and the National Defense Science \& Engineering Graduate (NDSEG) Fellowship Program. RCA also thanks NASA grants 80NSSC24K0908 and 80NSSC22K0993.  This work is supported by ERC grant (HELIO4CAST, 10.3030/101042188). Funded by the European Union. Views and opinions expressed are however those of the author(s) only and do not necessarily reflect those of the European Union or the European Research Council Executive Agency. Neither the European Union nor the granting authority can be held responsible for them. 
In-situ spacecraft data from Solar Orbiter, PSP, Wind, ACE, and STEREO-A are publicly available at NASA's Coordinated Data Analysis Web (CDAWeb) database \url{https://cdaweb.gsfc. nasa.gov/index.html/}. The ICMECAT version used in this study can be found at \url{https://doi.org/10.6084/m9.figshare.6356420.v22}.The Solar MACH tool is publicly available, and can be found at \url{https://solar-mach.streamlit.app/}. We acknowledge the Community Coordinated Modeling Center (CCMC) at Goddard Space Flight Center for the use of the DONKI catalog.

\end{acknowledgements}

%
%

\bibliographystyle{aa}
\bibliography{main}

\appendix 

\section{Calculation of interplanetary shock parameters}\label{app:shockcalc}

The calculation of the plasma parameters reported in Table~\ref{tab:parameters} was done using the methods employed by the Heliospheric Shock Database, generated and maintained at the University of Helsinki. Many of the equations used can be found in the accompanying paper, \citet{kil2015}, as well as the documentation provided on their website, \url{http://www.ipshocks.fi/documentation}. Many of the formulas used can also be found in Appendix A of \citet{walker2025}. Here we report the formulas not shown in that paper, specifically those used to calculate upstream magnetic turbulence, magnetic compression, and density compression.

As discussed in \citet{walker2025}, fixed time intervals for the upstream and downstream (denoted by the "u" and "d" subscripts, respectively) regions of the IP shock were used for each spacecraft. The upstream interval begins roughly 9 minutes before IP shock front measurement and ends 1 minute before the IP shock, while the interval begins 2 minutes after IP shock front measurement and extends to 10 minutes after the IP shock. We set the minimum number of data points within these intervals at three, though we allow them to extend if this minimum is not met. Average quantities for the upstream (downstream) intervals are denoted by the “u” (“d”) subscript, and upstream values (marked via the “up” superscript) are also averaged over the upstream time intervals.

The upstream magnetic turbulence is defined as the ratio between the fluctuations in the magnetic field ($\delta B$), which we calculate by finding the root-mean-square of the field, and the magnitude of the magnetic field ($B_0$). The equation is thus
\begin{equation}
    \frac{\delta B}{B_0} = \left\langle\frac{dB_{rms}}{B_0} \right\rangle^{up} = \frac{\sqrt{\langle(\mathbf{B_{u}} - \langle \mathbf{B_{u}} \rangle)^2\rangle}}{B_0}.
\end{equation}

The magnetic compression is defined as
\begin{equation}
    B_{comp} = \frac{B_{0,d}}{B_{0,u}}
\end{equation}
where $B_{0,u}$ ($B_{0,d}$)is the magnitude of the magnetic field over the upstream (downstream) time interval.

Similarly, the density compression is defined as 
\begin{equation}
    n_{comp} = \frac{n_d}{n_u}
\end{equation}
where $n_u$ ($n_d$) is the average plasma density over the upstream (downstream) time interval.

\section{Relevant Parameters for Multipoint Observations} \label{app:obstable}
\begin{sidewaystable*}[h]
    \centering
\small
    \caption{Local shock parameters and fitted spectral shape parameters for 23 events included in the analysis for this study. The shape parameters are only for observations of \textsuperscript{4}He. PS, CDPL, and PL refer to the Pan-spectrum fitting formula, classic double-power-law, and power law respectively.}
\begin{tabular}{lrllllllllll}
\toprule
Event & Spacecraft & ESP Start Time & ESP End Time & Best Fit Type & $E_0$ & $\beta_1$ & $\beta_2$ & $r$ (au) & $\theta_{Bn}$ & $\delta B/B_{0}$ & $B_{\mathrm{comp}}$ \\
\midrule
4 & STEREO A & 2023-11-13 08:30 & 2023-11-13 13:41 & PS & 0.172 $\pm$ 0.008 & 0.249 $\pm$ 0.248 & 3.99 $\pm$ 0.235 & 0.97 & 19.40 $\pm$ 5.69 & 0.16 $\pm$ 0.02 & 1.48 $\pm$ 0.13 \\
4 & Wind/ACE & 2023-11-12 04:00 & 2023-11-12 05:39 & PL & --- & 2.818 $\pm$ 0.032 & --- & 0.98 & 70.67 $\pm$ 7.51 & 0.16 $\pm$ 0.03 & 2.97 $\pm$ 0.31 \\
4 & Solar Orbiter & 2023-11-11 02:00 & 2023-11-11 09:13 & PS & 0.223 $\pm$ 0.002 & 0.554 $\pm$ 0.017 & 3.619 $\pm$ 0.021 & 0.69 & 38.84 $\pm$ 19.55 & 2.43 $\pm$ 0.35 & 0.91 $\pm$ 0.38 \\
6 & Wind/ACE & 2023-10-20 10:06 & 2023-10-20 12:33 & PS & 0.112 $\pm$ 0.001 & 1.041 $\pm$ 0.022 & 2.349 $\pm$ 0.021 & 0.99 & 60.63 $\pm$ 19.78 & 0.29 $\pm$ 0.06 & 1.63 $\pm$ 0.27 \\
6 & Solar Orbiter & 2023-10-17 09:31 & 2023-10-17 10:23 & PL & --- & 5.08 $\pm$ 0.071 & --- & 0.37 & 36.38 $\pm$ 16.34 & 2.69 $\pm$ 0.66 & 1.73 $\pm$ 1.08 \\
7 & Wind/ACE & 2023-09-24 14:00 & 2023-09-24 20:01 & PS & 0.116 $\pm$ 0.0 & 1.139 $\pm$ 0.0 & 1.802 $\pm$ 0.0 & 0.99 & 17.97 $\pm$ 13.55 & 0.25 $\pm$ 0.04 & 2.44 $\pm$ 0.50 \\
7 & STEREO A & 2023-09-24 09:00 & 2023-09-24 17:38 & PS & 0.195 $\pm$ 0.003 & 0.693 $\pm$ 0.078 & 2.845 $\pm$ 0.008 & 0.96 & 29.70 $\pm$ 21.04 & 0.57 $\pm$ 0.12 & 1.91 $\pm$ 1.16 \\
7 & PSP & 2023-09-22 07:00 & 2023-09-22 18:14 & PL & --- & 2.532 $\pm$ 0.159 & --- & 0.25 & 26.62 $\pm$ 19.53 & 0.33 $\pm$ 0.07 & 1.47 $\pm$ 0.15 \\
9 & Wind/ACE & 2023-09-18 06:00 & 2023-09-18 12:51 & PS & 0.098 $\pm$ 0.0 & 0.499 $\pm$ 0.0 & 1.827 $\pm$ 0.0 & 1.00 & 79.36 $\pm$ 4.77 & 0.08 $\pm$ 0.01 & 2.40 $\pm$ 0.14 \\
9 & STEREO A & 2023-09-18 00:00 & 2023-09-18 09:03 & PS & 0.193 $\pm$ 0.001 & 0.832 $\pm$ 0.011 & 3.04 $\pm$ 0.005 & 0.96 & 51.54 $\pm$ 10.01 & 0.33 $\pm$ 0.06 & 2.29 $\pm$ 0.16 \\
9 & PSP & 2023-09-16 18:00 & 2023-09-17 02:36 & CDPL & 0.044 $\pm$ 0.012 & 3.829 $\pm$ 0.546 & 4.222 $\pm$ 0.205 & 0.41 & 33.11 $\pm$ 17.48 & 0.89 $\pm$ 0.21 & 2.03 $\pm$ 0.92 \\
10 & Wind/ACE & 2023-09-12 02:00 & 2023-09-12 10:23 & PS & 0.076 $\pm$ 0.001 & 0.114 $\pm$ 0.103 & 2.657 $\pm$ 0.028 & 1.00 & 15.60 $\pm$ 9.28 & 0.40 $\pm$ 0.07 & 1.26 $\pm$ 0.23 \\
10 & STEREO A & 2023-09-11 21:00 & 2023-09-12 07:04 & --- & --- & --- & --- & 0.96 & 29.53 $\pm$ 14.18 & 0.24 $\pm$ 0.03 & 1.97 $\pm$ 0.27 \\
10 & PSP & 2023-09-10 01:00 & 2023-09-10 10:04 & PL & --- & 7.598 $\pm$ 0.0 & --- & 0.54 & 18.78 $\pm$ 8.07 & 0.19 $\pm$ 0.03 & 1.34 $\pm$ 0.14 \\
13 & PSP & 2023-07-16 01:47 & 2023-07-16 02:32 & --- & --- & --- & --- & 0.63 & 58.12 $\pm$ 21.41 & 0.54 $\pm$ 0.11 & 1.87 $\pm$ 0.53 \\
13 & STEREO A & 2023-07-16 21:06 & 2023-07-16 22:33 & PS & 0.19 $\pm$ 0.006 & 1.49 $\pm$ 0.237 & 4.346 $\pm$ 0.043 & 0.96 & 38.10 $\pm$ 21.62 & 0.94 $\pm$ 0.31 & 1.25 $\pm$ 0.48 \\
14 & PSP & 2023-04-23 03:43 & 2023-04-23 06:10 & --- & --- & --- & --- & 0.73 & 74.13 $\pm$ 11.69 & 0.22 $\pm$ 0.03 & 1.95 $\pm$ 0.24 \\
14 & Solar Orbiter & 2023-04-22 06:56 & 2023-04-22 08:01 & PS & 0.188 $\pm$ 0.009 & 0.758 $\pm$ 0.137 & 4.258 $\pm$ 0.12 & 0.39 & 48.24 $\pm$ 16.26 & 1.74 $\pm$ 0.16 & 1.97 $\pm$ 0.21 \\
15 & STEREO A & 2023-03-21 23:00 & 2023-03-22 05:52 & CDPL & 0.193 $\pm$ 0.026 & 1.65 $\pm$ 0.4 & 3.224 $\pm$ 0.154 & 0.97 & 29.51 $\pm$ 15.37 & 1.12 $\pm$ 0.39 & 1.70 $\pm$ 0.59 \\
15 & Wind/ACE & 2023-03-23 00:00 & 2023-03-23 05:53 & CDPL & 0.106 $\pm$ 0.002 & 1.568 $\pm$ 0.029 & 3.058 $\pm$ 0.089 & 0.99 & 64.67 $\pm$ 20.09 & 0.20 $\pm$ 0.03 & 1.09 $\pm$ 0.17 \\
15 & Solar Orbiter & 2023-03-21 08:30 & 2023-03-21 12:22 & PS & 0.256 $\pm$ 0.012 & 2.252 $\pm$ 0.212 & 4.295 $\pm$ 0.079 & 0.50 & 54.66 $\pm$ 15.96 & 1.46 $\pm$ 0.10 & 0.76 $\pm$ 0.11 \\
16 & Solar Orbiter & 2023-03-14 00:30 & 2023-03-14 01:07 & PS & 0.356 $\pm$ 0.035 & 1.739 $\pm$ 0.064 & 3.926 $\pm$ 0.178 & 0.60 & 62.47 $\pm$ 17.19 & 3.47 $\pm$ 1.02 & 3.47 $\pm$ 1.22 \\
16 & STEREO A & 2023-03-14 22:00 & 2023-03-15 01:15 & PS & 0.282 $\pm$ 0.001 & 1.213 $\pm$ 0.001 & 2.605 $\pm$ 0.004 & 0.97 & 34.35 $\pm$ 22.07 & 0.35 $\pm$ 0.04 & 1.99 $\pm$ 0.80 \\
17 & Solar Orbiter & 2023-03-09 03:42 & 2023-03-09 05:57 & PL & --- & --- & --- & 0.65 & 56.22 $\pm$ 21.54 & 2.39 $\pm$ 0.39 & 1.07 $\pm$ 0.27 \\
17 & STEREO A & 2023-03-10 10:27 & 2023-03-10 11:50 & PL & --- & 1.337 $\pm$ 0.237 & --- & 0.97 & 19.55 $\pm$ 10.73 & 0.59 $\pm$ 0.12 & 1.43 $\pm$ 0.26 \\
18 & Solar Orbiter & 2023-03-08 17:43 & 2023-03-08 20:34 & PL & --- & 3.89 $\pm$ 0.241 & --- & 0.66 & 61.99 $\pm$ 12.67 & 0.53 $\pm$ 0.03 & 0.67 $\pm$ 0.06 \\
18 & STEREO A & 2023-03-09 16:02 & 2023-03-09 19:14 & PL & --- & 1.504 $\pm$ 0.0 & --- & 0.97 & 32.45 $\pm$ 22.12 & 0.78 $\pm$ 0.27 & 1.40 $\pm$ 0.51 \\
20 & Solar Orbiter & 2022-09-06 06:48 & 2022-09-06 10:00 & PS & 1.304 $\pm$ 0.022 & 0.371 $\pm$ 0.008 & 1.211 $\pm$ 0.012 & 0.69 & 51.27 $\pm$ 21.00 & 2.58 $\pm$ 0.64 & 1.44 $\pm$ 0.77 \\
20 & PSP & 2022-09-05 16:24 & 2022-09-05 17:26 & CDPL & 0.081 $\pm$ 0.003 & 1.922 $\pm$ 0.094 & 4.405 $\pm$ 0.197 & 0.07 & 57.80 $\pm$ 16.28 & 0.32 $\pm$ 0.06 & 2.34 $\pm$ 0.21 \\
21 & STEREO A & 2022-08-19 19:46 & 2022-08-19 20:56 & CDPL & 0.195 $\pm$ 0.017 & 2.463 $\pm$ 0.215 & 4.476 $\pm$ 0.289 & 0.96 & 20.93 $\pm$ 10.19 & 0.37 $\pm$ 0.05 & 1.94 $\pm$ 0.63 \\
21 & PSP & 2022-08-18 11:50 & 2022-08-18 13:27 & PL & --- & 5.389 $\pm$ 0.486 & --- & 0.55 & 29.54 $\pm$ 22.03 & 0.21 $\pm$ 0.02 & 1.32 $\pm$ 0.66 \\
22 & PSP & 2022-07-17 12:44 & 2022-07-17 13:31 & --- & --- & --- & --- & 0.76 & 59.93 $\pm$ 17.35 & 0.26 $\pm$ 0.05 & 1.32 $\pm$ 0.06 \\
22 & Wind/ACE & 2022-07-18 19:21 & 2022-07-18 20:25 & PS & 0.077 $\pm$ 0.0 & 0.043 $\pm$ 0.038 & 4.547 $\pm$ 0.03 & 1.00 & 63.00 $\pm$ 8.46 & 0.23 $\pm$ 0.04 & 1.57 $\pm$ 0.22 \\
23 & STEREO A & 2022-03-12 15:30 & 2022-03-12 21:12 & CDPL & 0.202 $\pm$ 0.01 & 1.944 $\pm$ 0.158 & 3.766 $\pm$ 0.128 & 0.97 & 33.32 $\pm$ 22.71 & 0.77 $\pm$ 0.22 & 1.32 $\pm$ 0.43 \\
23 & Wind/ACE & 2022-03-13 01:30 & 2022-03-13 10:03 & PS & 0.919 $\pm$ 0.296 & 0.548 $\pm$ 0.167 & 9.309 $\pm$ 1.968 & 0.98 & 46.72 $\pm$ 21.78 & 0.86 $\pm$ 0.33 & 1.02 $\pm$ 1.60 \\
23 & Solar Orbiter & 2022-03-11 16:02 & 2022-03-11 19:52 & PS & 0.185 $\pm$ 0.016 & 2.64 $\pm$ 0.222 & 3.349 $\pm$ 0.021 & 0.44 & 30.44 $\pm$ 12.25 & 1.52 $\pm$ 0.18 & 1.59 $\pm$ 0.53 \\
24 & Wind/ACE & 2022-03-10 09:00 & 2022-03-10 16:11 & PL & --- & 2.786 $\pm$ 0.302 & --- & 0.98 & 80.36 $\pm$ 9.15 & 0.09 $\pm$ 0.02 & 1.29 $\pm$ 0.25 \\
24 & Solar Orbiter & 2022-03-08 05:00 & 2022-03-08 14:45 & CDPL & 0.261 $\pm$ 0.018 & 3.955 $\pm$ 0.097 & 4.868 $\pm$ 0.146 & 0.48 & 51.30 $\pm$ 10.85 & 0.63 $\pm$ 0.03 & 1.44 $\pm$ 0.10 \\
25 & STEREO A & 2021-11-11 19:00 & 2021-11-11 23:10 & CDPL & 0.482 $\pm$ 0.103 & 3.08 $\pm$ 0.326 & 0.335 $\pm$ 0.657 & 0.96 & 27.30 $\pm$ 13.16 & 0.37 $\pm$ 0.05 & 2.42 $\pm$ 0.32 \\
25 & PSP & 2021-11-09 11:30 & 2021-11-09 16:21 & --- & --- & --- & --- & 0.42 & 77.67 $\pm$ 5.68 & 0.07 $\pm$ 0.01 & 2.07 $\pm$ 0.15 \\
28 & Solar Orbiter & 2021-10-14 15:00 & 2021-10-14 23:11 & PL & --- & 2.083 $\pm$ 0.107 & --- & 0.72 & 39.15 $\pm$ 20.16 & 0.89 $\pm$ 0.06 & 1.52 $\pm$ 0.13 \\
28 & Wind/ACE & 2021-10-15 19:00 & 2021-10-16 00:07 & PL & --- & 0.63 $\pm$ 1.152 & --- & 0.99 & 40.92 $\pm$ 22.14 & 0.67 $\pm$ 0.26 & 1.76 $\pm$ 0.58 \\
29 & Solar Orbiter & 2021-10-03 23:22 & 2021-10-04 00:30 & PL & --- & 5.703 $\pm$ 0.0 & --- & 0.65 & 17.36 $\pm$ 10.17 & 12.65 $\pm$ 2.16 & 1.96 $\pm$ 0.36 \\
29 & Wind/ACE & 2021-10-05 23:58 & 2021-10-06 02:19 & --- & --- & --- & --- & 0.99 & 56.84 $\pm$ 14.43 & 0.06 $\pm$ 0.01 & 1.25 $\pm$ 0.04 \\
30 & PSP & 2021-09-26 07:10 & 2021-09-26 08:54 & --- & --- & --- & --- & 0.78 & 70.59 $\pm$ 13.74 & 0.78 $\pm$ 0.16 & 3.00 $\pm$ 0.59 \\
30 & Wind/ACE & 2021-09-26 22:52 & 2021-09-27 00:19 & --- & --- & --- & --- & 0.99 & 19.91 $\pm$ 12.14 & 0.26 $\pm$ 0.04 & 1.29 $\pm$ 0.17 \\
31 & STEREO A & 2021-09-13 03:00 & 2021-09-13 16:57 & CDPL & 0.442 $\pm$ 0.261 & 2.815 $\pm$ 1.26 & -0.231 $\pm$ 1.144 & 0.96 & 36.49 $\pm$ 8.49 & 0.54 $\pm$ 0.12 & 1.17 $\pm$ 0.25 \\
31 & PSP & 2021-09-12 03:00 & 2021-09-12 15:52 & --- & --- & --- & --- & 0.72 & 30.51 $\pm$ 21.35 & 0.25 $\pm$ 0.03 & 0.96 $\pm$ 0.13 \\
33 & STEREO A & 2021-05-04 12:15 & 2021-05-04 13:24 & --- & --- & --- & --- & 0.97 & 63.14 $\pm$ 15.56 & 0.25 $\pm$ 0.04 & 1.56 $\pm$ 0.10 \\
33 & PSP & 2021-04-30 05:07 & 2021-04-30 05:53 & --- & --- & --- & --- & 0.09 & 41.08 $\pm$ 22.17 & 0.22 $\pm$ 0.03 & 1.32 $\pm$ 0.16 \\
35 & PSP & 2020-06-25 10:51 & 2020-06-25 11:38 & --- & --- & --- & --- & 0.52 & 11.90 $\pm$ 6.10 & 0.13 $\pm$ 0.02 & 1.33 $\pm$ 0.08 \\
35 & Wind/ACE & 2020-06-29 23:25 & 2020-06-30 01:11 & --- & --- & --- & --- & 1.01 & 69.58 $\pm$ 16.03 & 0.87 $\pm$ 0.46 & 1.60 $\pm$ 0.81 \\
\bottomrule
\end{tabular}
\label{tab:parameters}
\end{sidewaystable*}

\begin{center}
\small
\tablefirsthead{%
    \toprule
    Event & Spacecraft & Shock Time & $r$ (au) & $Long$ (deg)\\
    \midrule}
\tablehead{%
    \multicolumn{5}{l}{\small\textit{Table \thetable{} continued}} \\
    \toprule
    Event & Spacecraft & Shock Time & $r$ (au)& $Long$ (deg)\\
    \midrule}
\tabletail{%
    \midrule
    \multicolumn{5}{r}{\small\textit{Continued}} \\}
\tablelasttail{\bottomrule}

\bottomcaption{Full catalog of 39 multipoint ICME events compiled for this study. Longitude is given in HEEQ coordinates.}
\begin{supertabular}{lrlll}
1 & STA & 2023-12-15 12:19 & 0.97 & 6.70 \\
1 & Wind & 2023-12-15 10:52 & 0.97 & -0.18 \\
1 & SolO & 2023-12-13 07:44 & 0.91 & -12.04 \\
1 & SOHO & 2023-12-13 00:00 & 0.98 & --- \\
2 & Wind & 2023-12-01 08:52 & 0.98 & -0.24 \\
2 & STA & 2023-12-01 08:24 & 0.97 & 6.46 \\
2 & SolO & 2023-12-01 02:27 & 0.85 & -10.16 \\
2 & SOHO & 2023-11-28 20:24 & 0.98 & --- \\
2 & STA/HIA & 2023-11-28 22:47 & 0.97 & --- \\
3 & Wind & 2023-11-30 23:26 & 0.98 & -0.24 \\
3 & SolO & 2023-11-30 10:47 & 0.84 & -10.13 \\
3 & SOHO & 2023-11-27 06:48 & 0.98 & --- \\
4 & STA & 2023-11-13 13:42 & 0.97 & 6.03 \\
4 & Wind & 2023-11-12 05:40 & 0.98 & -0.23 \\
4 & SolO & 2023-11-11 09:14 & 0.69 & -12.33 \\
4 & STA/COR2 & 2023-11-09 12:23 & 0.96 & --- \\
4 & STA/HIA & 2023-11-09 16:48 & 0.96 & --- \\
5 & Wind & 2023-11-04 13:34 & 0.98 & -0.19 \\
5 & STA & 2023-11-04 00:20 & 0.96 & 5.73 \\
6 & Wind & 2023-10-20 12:34 & 0.99 & -0.08 \\
6 & SolO & 2023-10-17 10:24 & 0.37 & -51.15 \\
6 & STA/COR2 & 2023-10-16 12:09 & 0.96 & --- \\
6 & STA/HIA & 2023-10-16 15:28 & 0.96 & --- \\
7 & Wind & 2023-09-24 20:02 & 0.99 & 0.15 \\
7 & STA & 2023-09-24 17:39 & 0.96 & 3.68 \\
7 & PSP & 2023-09-22 18:15 & 0.25 & 14.18 \\
8 & Wind & 2023-09-19 14:30 & 1.00 & 0.19 \\
8 & STA & 2023-09-19 00:52 & 0.96 & 3.25 \\
8 & PSP & 2023-09-17 13:52 & 0.40 & 3.64 \\
8 & SOHO & 2023-09-16 09:12 & 0.99 & --- \\
9 & Wind & 2023-09-18 12:52 & 1.00 & 0.19 \\
9 & STA & 2023-09-18 09:04 & 0.96 & 3.21 \\
9 & PSP & 2023-09-17 02:37 & 0.41 & 3.31 \\
9 & SOHO & 2023-09-14 23:12  & 0.99 & --- \\
10 & Wind & 2023-09-12 10:24 & 1.00 & 0.23 \\
10 & STA & 2023-09-12 07:05 & 0.96 & 2.75 \\
10 & PSP & 2023-09-10 10:05 & 0.54 & 0.58 \\
10 & SOHO & 2023-09-07 02:12 & 1.00 & --- \\
10 & STA/HIA & 2023-09-07 02:26 & 0.96 & --- \\
11 & Wind & 2023-08-05 02:04 & 1.00 & 0.17 \\
11 & STA & 2023-08-04 22:55 & 0.96 & -0.70 \\
11 & STA/HIA & 2023-08-02 10:01 & 0.96 & --- \\
12 & Wind & 2023-08-01 11:12 & 1.00 & 0.15 \\
12 & STA & 2023-08-01 09:23 & 0.96 & -1.01 \\
12 & STA/HIA & 2023-07-27 20:21 & 0.96 & --- \\
13 & STA & 2023-07-16 22:34 & 0.96 & -2.60 \\
13 & PSP & 2023-07-16 02:33 & 0.63 & 16.94 \\
13 & SOHO & 2023-07-14 15:16 & 1.01 & --- \\
13 & STA/HIA & 2023-07-14 17:28 & 0.96 & --- \\
14 & PSP & 2023-04-23 06:11 & 0.73 & 105.86 \\
14 & SolO & 2023-04-22 08:02 & 0.39 & 127.26 \\
14 & SOHO & 2023-04-21 06:24 & 0.99 & --- \\
15 & Wind & 2023-03-23 05:54 & 0.99 & 0.19 \\
15 & STA & 2023-03-22 05:53 & 0.97 & -11.97 \\
15 & SolO & 2023-03-21 12:23 & 0.50 & -16.79 \\
15 & SOHO & 2023-03-20 14:42 & 0.98 & --- \\
16 & STA & 2023-03-15 01:16 & 0.97 & -12.29 \\
16 & SolO & 2023-03-14 01:08 & 0.60 & -25.74 \\
16 & SOHO & 2023-03-12 23:37 & 0.98 & --- \\
17 & STA & 2023-03-10 11:51 & 0.97 & -12.47 \\
17 & SolO & 2023-03-09 05:58 & 0.65 & -29.07 \\
18 & STA & 2023-03-09 19:15 & 0.97 & -12.48 \\
18 & SolO & 2023-03-08 20:35 & 0.66 & -29.27 \\
19 & STA & 2022-12-26 07:51 & 0.97 & -13.79 \\
19 & SolO & 2022-12-26 05:21 & 0.93 & -19.64 \\
19 & SOHO & 2022-12-24 02:48 & 0.97 & --- \\
20 & SolO & 2022-09-06 10:01 & 0.69 & 150.00 \\
20 & PSP & 2022-09-05 17:27 & 0.07 & 124.67 \\
20 & STA/HIA & 2022-09-05 18:09 & 0.96 & --- \\
20 & SOHO & 2022-09-05 16:36 & 1.00 & --- \\
21 & STA & 2022-08-19 20:57 & 0.96 & -21.00 \\
21 & PSP & 2022-08-18 13:28 & 0.55 & 21.41 \\
21 & STA/HIA & 2022-08-15 22:07 & 0.96 & --- \\
21 & SOHO & 2022-08-17 14:53 & 1.00 & --- \\
22 & Wind & 2022-07-18 20:26 & 1.00 & -0.04 \\
22 & PSP & 2022-07-17 13:32 & 0.76 & 29.62 \\
22 & STA/HIA & 2022-07-15 13:54 & 0.96 & --- \\
23 & Wind & 2022-03-13 10:04 & 0.98 & 0.25 \\
23 & STA & 2022-03-12 21:13 & 0.97 & -33.86 \\
23 & SolO & 2022-03-11 19:53 & 0.44 & 9.96 \\
23 & STA/COR2 & 2022-03-10 19:23 & 0.97 & --- \\
23 & STA/HIA & 2022-03-10 18:14 & 0.97 & --- \\
24 & Wind & 2022-03-10 16:12 & 0.98 & 0.25 \\
24 & SolO & 2022-03-08 14:46 & 0.48 & 2.95 \\
24 & SOHO & 2022-03-07 00:12 & 0.98 & --- \\
24 & STA/HIA & 2022-03-06 22:48 & 0.97 & --- \\
25 & STA & 2021-11-11 23:11 & 0.96 & -36.73 \\
25 & PSP & 2021-11-09 16:22 & 0.42 & -50.99 \\
25 & SOHO & 2021-11-07 15:05 & 0.98 & --- \\
26 & STA & 2021-11-04 04:15 & 0.96 & -37.20 \\
26 & Wind & 2021-11-03 19:40 & 0.98 & -0.04 \\
26 & SolO & 2021-11-03 14:03 & 0.85 & -0.96 \\
26 & STA/COR2 & 2021-11-01 18:38 & 0.96 & --- \\
26 & SOHO & 2021-11-01 21:36 & 0.98 & --- \\
26 & SOHO & 2021-11-02 02:48 & 0.98 & --- \\
26 & STA/HIA & 2021-11-02 01:30 & 0.96 & --- \\
26 & STA/HIA & 2021-10-31 20:56 & 0.96 & --- \\
27 & Wind & 2021-10-31 09:32 & 0.98 & -0.01 \\
27 & STA & 2021-10-31 07:57 & 0.99 & -37.42 \\
27 & SolO & 2021-10-30 22:01 & 0.82 & -2.01 \\
27 & STA/COR2 & 2021-10-28 15:53 & 0.96 & --- \\
27 & STA/COR2 & 2021-10-28 01:23 & 0.96 & --- \\
27 & STA/HIA & 2021-10-28 10:25 & 0.96 & --- \\
28 & Wind & 2021-10-16 00:08 & 0.99 & 0.13 \\
28 & SolO & 2021-10-14 23:12 & 0.72 & -9.79 \\
28 & SOHO & 2021-10-12 03:24 & 0.99 & --- \\
28 & STA/HIA & 2021-10-11 22:59 & 0.96 & --- \\
29 & Wind & 2021-10-06 02:20 & 0.99 & 0.20 \\
29 & SolO & 2021-10-04 00:31 & 0.65 & -19.72 \\
29 & SOHO & 2021-10-01 16:12 & 0.99 & --- \\
29 & STA/HIA & 2021-10-01 17:38 & 0.96 & --- \\
30 & Wind & 2021-09-27 00:20 & 0.99 & 0.24 \\
30 & PSP & 2021-09-26 08:55 & 0.78 & -43.07 \\
30 & SOHO & 2021-09-23 16:53 & 0.99 & --- \\
31 & STA & 2021-09-13 16:58 & 0.96 & -40.85 \\
31 & PSP & 2021-09-12 15:53 & 0.72 & -38.06 \\
32 & Wind & 2021-05-26 11:44 & 1.00 & -0.19 \\
32 & STA & 2021-05-26 06:11 & 0.96 & -50.64 \\
33 & STA & 2021-05-04 13:25 & 0.97 & -52.47 \\
33 & PSP & 2021-04-30 05:54 & 0.09 & -26.32 \\
33 & SOHO & 2021-04-29 19:24 & 1.00 & --- \\
33 & STA/HIA & 2021-04-29 22:41 & 0.97 & --- \\
34 & STA & 2020-12-01 07:28 & 0.96 & -57.60 \\
34 & PSP & 2020-12-01 02:22 & 0.80 & -96.85 \\
34 & STA/COR2 & 2020-11-29 13:24 & 0.96 & --- \\
35 & Wind & 2020-06-30 01:12 & 1.01 & -0.25 \\
35 & PSP & 2020-06-25 11:39 & 0.52 & 19.91 \\
35 & STA/COR2 & 2020-06-22 15:09 & 0.96 & --- \\
35 & STA/HIA & 2020-06-22 18:33 & 0.96 & --- \\
36 & Wind & 2020-04-20 01:34 & 1.00 & 0.18 \\
36 & SolO & 2020-04-19 05:06 & 0.81 & -3.98 \\
36 & STA/COR2 & 2020-04-14 21:54 & 0.97 & --- \\
36 & STA/HIA & 2020-04-15 06:28 & 0.97 & --- \\
37 & STA & 2019-10-14 07:44 & 0.96 & -82.60 \\
37 & PSP & 2019-10-13 19:03 & 0.81 & -74.94 \\
37 & SOHO & 2019-10-10 11:05 & 0.99 & --- \\
38 & MAVEN & 2016-07-21 15:48 & 1.45 & -24.03 \\
38 & Wind & 2016-07-19 23:01 & 1.01 & -0.21 \\
38 & SOHO & 2016-07-17 12:48 & 1.01 & --- \\
38 & STA/HIA & 2016-07-17 07:58 & 0.97 & --- \\
39 & MAVEN & 2016-01-19 01:47 & 1.65 & 64.35 \\
39 & Wind & 2016-01-18 21:18 & 0.98 & -0.20 \\
39 & SOHO & 2016-01-15 00:00 & 0.97 & --- \\

\end{supertabular}
\label{tab:fullcatalog}
\end{center}

\section{Dependence of Spectral Break Energy on Shock Theta for a Selection of Ion Species}\label{app:thetaunc}
\begin{figure*}
    \centering
    \includegraphics[width=1\linewidth]{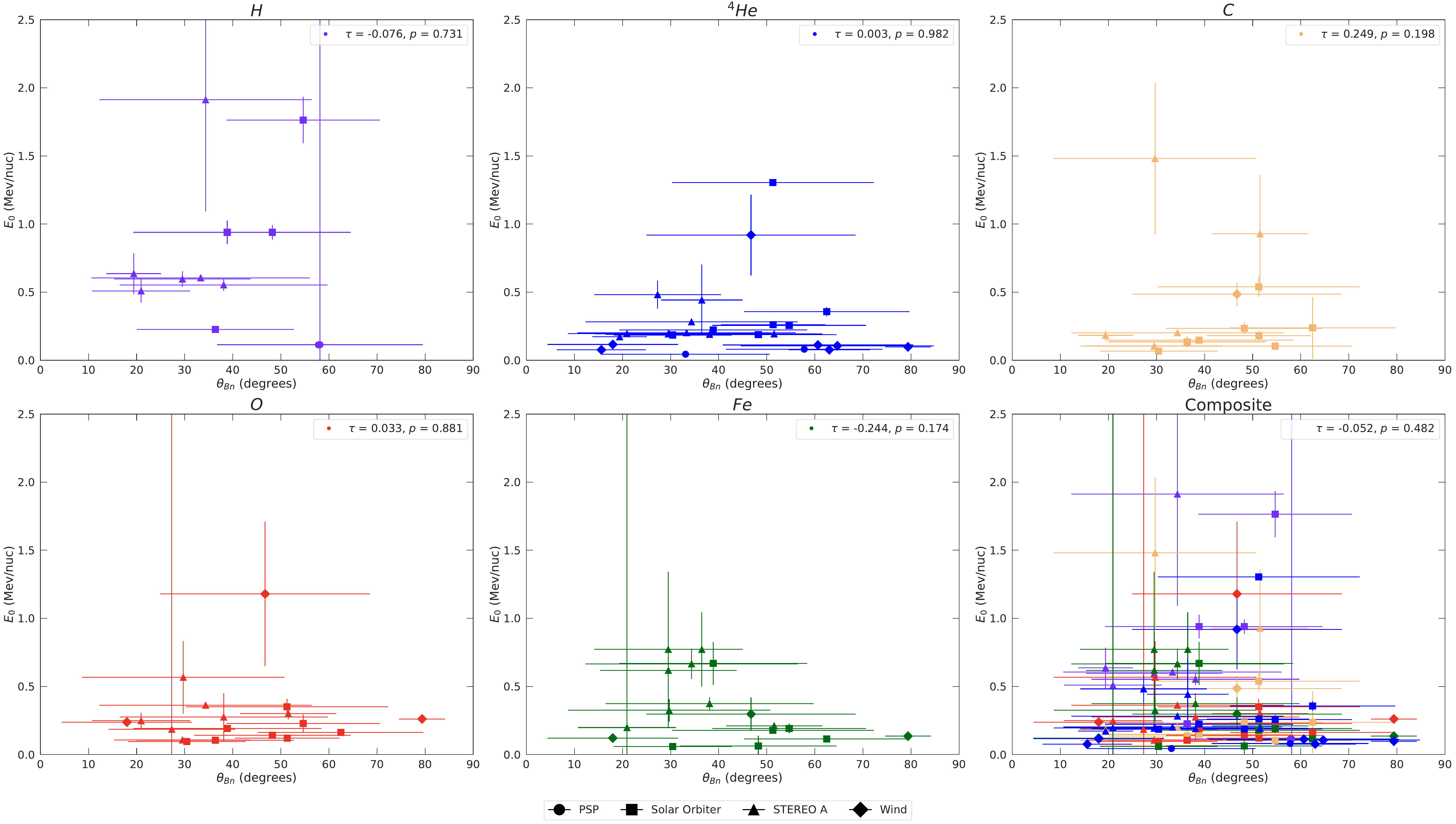}
    \caption{Scatter plots showing the relationship between break energy ($E_0$) and shock normal angle ($\theta_{Bn}$) for different ion species measured at individual spacecraft for the 23 multipoint events used for this study. The calculated Kendell's tau coefficient ($\tau$) and p-values ($p$) for these species are shown both in their respective panels, and in panel e, which depicts observations of all species with corresponding uncertainties as an extension of Figure \ref{fig:kendelltau}. }
    \label{fig:fulltau}
\end{figure*}
\end{document}